\documentclass{article}

\usepackage{microtype}
\usepackage{graphicx}
\usepackage{subcaption}
\usepackage{booktabs} 

\usepackage{hyperref}

 \usepackage[preprint]{icml2026}

\usepackage{amsmath}
\usepackage{amssymb}
\usepackage{mathtools}
\usepackage{amsthm}

\usepackage[capitalize,noabbrev]{cleveref}

\usepackage{multirow}  
\usepackage{listings}
\usepackage{graphicx}
 
\usepackage{colortbl}

\definecolor{Wheat1}{RGB}{255, 248, 200}
\definecolor{Wheat2}{RGB}{255, 222, 115}

\definecolor{Grey1}{RGB}{235,235,235}
\definecolor{Grey2}{RGB}{215,215,215}
\definecolor{White}{RGB}{255,255,255}

\lstdefinelanguage{json}{
  basicstyle=\ttfamily\small,
  showstringspaces=false,
  breaklines=true,
  frame=single,
  backgroundcolor=\color{gray!5},
  literate=
   *{0}{{\textcolor{black}{0}}}{1}
    {1}{{\textcolor{black}{1}}}{1}
    {2}{{\textcolor{black}{2}}}{1}
    {3}{{\textcolor{black}{3}}}{1}
    {4}{{\textcolor{black}{4}}}{1}
    {5}{{\textcolor{black}{5}}}{1}
    {6}{{\textcolor{black}{6}}}{1}
    {7}{{\textcolor{black}{7}}}{1}
    {8}{{\textcolor{black}{8}}}{1}
    {9}{{\textcolor{black}{9}}}{1}
    {:}{{\textcolor{black}{:}}}{1}
    {,}{{\textcolor{black}{,}}}{1}
    {"}{{\textcolor{black}{"}}}{1},
}

\theoremstyle{plain}

\theoremstyle{definition}

\theoremstyle{remark}

\usepackage[textsize=tiny]{todonotes}

\icmltitlerunning{Predictive AI with External Knowledge Infusion}

\begin{document}

\twocolumn[

   \icmltitle{Predictive AI with External Knowledge Infusion: \\Datasets and Benchmarks for Stock Markets}



  \icmlsetsymbol{equal}{*}

  \begin{icmlauthorlist}
    \icmlauthor{Ambedkar Dukkipati}{yyy}
    \icmlauthor{Kawin Mayilvaghanan}{yyy}
    \icmlauthor{Naveen Kumar Pallekonda}{yyy}
    \icmlauthor{Sai Prakash Hadnoor}{yyy}
    \icmlauthor{Ranga Shaarad Ayyagari}{yyy}
  \end{icmlauthorlist}

  \icmlaffiliation{yyy}{Department of Computer Science and Automation, Indian Institute of Science, Bengaluru, INDIA}

  \icmlcorrespondingauthor{Ambedkar Dukkipati}{ambedkar@iisc.ac.in}


  \vskip 0.3in
]


\printAffiliationsAndNotice{}  

\begin{abstract}
Fluctuations in stock prices are influenced by a complex interplay of factors that go beyond mere historical data. These factors, themselves influenced by external forces, encompass inter-stock dynamics, broader economic factors, various government policy decisions, outbreaks of wars, etc. Furthermore, all of these factors are dynamic and exhibit changes over time. In this paper, for the first time, we tackle the forecasting problem under external influence by proposing learning mechanisms that not only learn from historical trends but also incorporate external knowledge from temporal knowledge graphs. Since there are no such datasets or temporal knowledge graphs available, we study this problem with stock market data, and we construct comprehensive temporal knowledge graph datasets. In our proposed approach, we model relations on external temporal knowledge graphs as events of a Hawkes process on graphs. With extensive experiments, we show that learned dynamic representations effectively rank stocks based on returns across multiple holding periods, outperforming related baselines on relevant metrics.

\end{abstract}

\section{Introduction} 
The development of learning algorithms for forecasting events is an important component in building AI systems with a wide range of applications, including in financial markets and healthcare. 
\begin{figure}[ht]
  \vskip 0.2in
  \begin{center}
    \centerline{\includegraphics[width=\columnwidth]{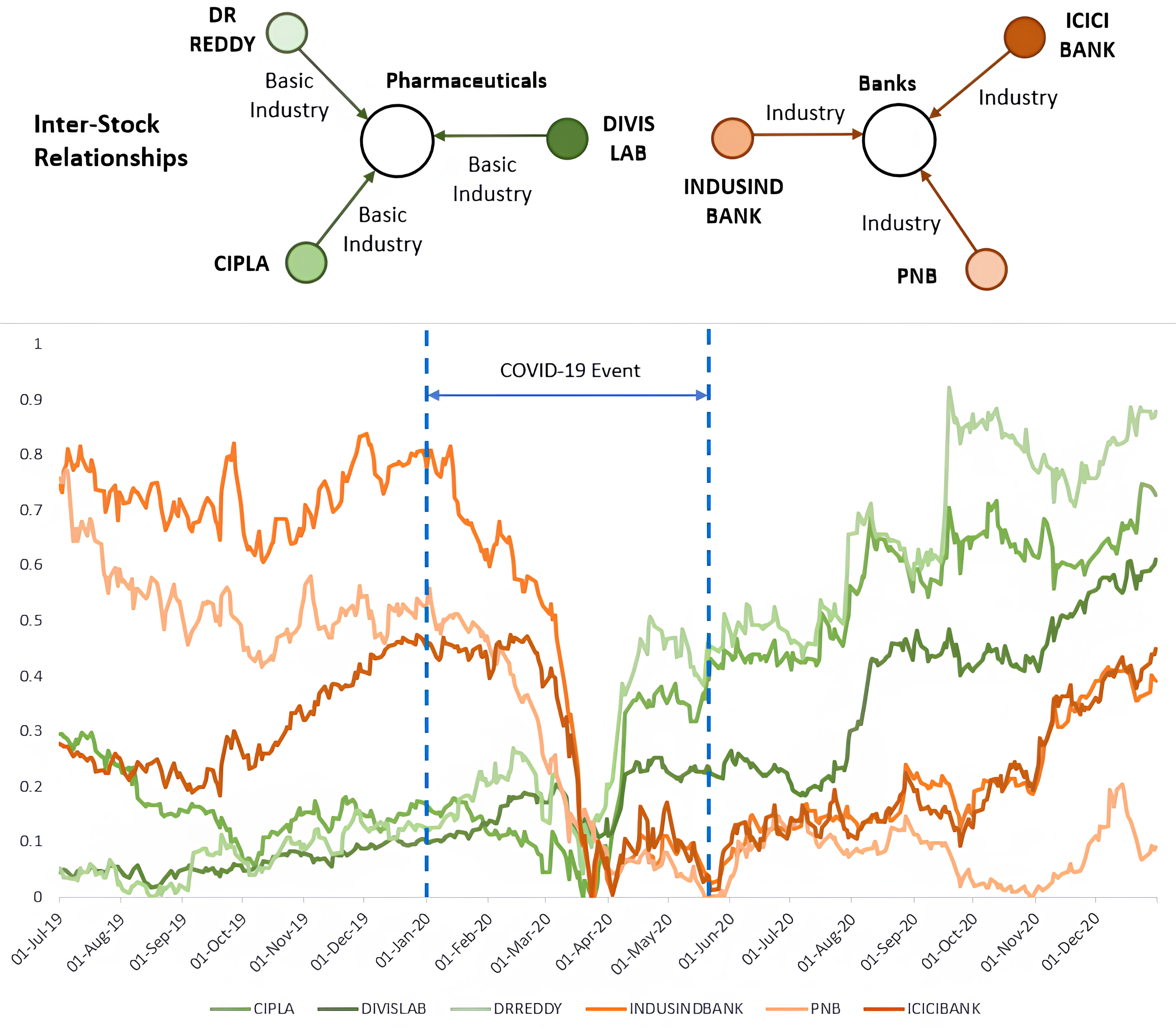}}
    \caption{
      Interdependence of stock trends between stocks in the same class during COVID-19 (January–May 2020, NSE). Pharmaceutical stocks (CIPLA, DIVISLAB, DRREDDY) trended positively, whereas banking stocks (PNB, INDUSINDBANK, ICICIBANK) declined.
    }
    \label{stock_dependence}
  \end{center}
\end{figure}
\vspace{-8pt}

Recent progress in these areas has been notable, and methods involving a combination of power sequence models, such as transformers, and elegant probabilistic models, such as temporal point processes, have made enormous progress~\cite{mei2017neuralhawkes,2021:GraciousGuptaCastroDukkipati:NeuralLatentSpaceModelForDynamicNetworks,2023:AAAI:GraciousDukkipati:HigherOrderInteractionForecasting,2025:AAAI:GraciousGuptaDukkipati:NeuralTemporalPointProcess,2025:AAAI:GraciousDukkipati:ForecastingRecursiveTemporalNetworks}. However, nearly all existing methods fall short of adequately accounting for or integrating time-varying external factors that influence the outcomes. For instance, stock price variations are affected by a complex interplay of factors that extend beyond mere historical patterns of stock prices. These elements can be shaped by external influences, such as prevailing economic conditions, shifts in governmental policies, and incidents of conflict. Additionally, these factors can be dynamic, with some changing rapidly, others very slowly, some very frequently, and others infrequently.



To study this problem, we develop an approach that captures external influences via temporal Knowledge Graphs (KG) and learns historical patterns via temporal point processes. Considering that such specific temporal KGs are not available in the literature, we develop temporal KGs for financial markets for the machine learning community and propose representation learning techniques to establish benchmark results. The temporal KGs we constructed integrated diverse components, including first- and second-order stock relationships, corporate events, macroeconomic indicators, events derived from financial statements, financial ratios disclosed during earnings calls, and analyst sentiments. Notably, all these elements are equipped with time-aware attributes. The capacity to incorporate entities of varying categories, such as Person, Stock Assets, Products, and Alliance, results in an intrinsically heterogeneous knowledge graph. We refer to these KGs as
$\mathsf{STOCKnowledge}$ aims to encapsulate all pertinent information related to stock assets within a given share market. 

\noindent
\textbf{Contributions}\\
\noindent
\textbf{(1)} We propose a time-series forecasting algorithm that incorporates knowledge of external events influencing the evolution of the series.\\
\noindent
    \textbf{(2)} For the task of stock prediction, we construct temporal knowledge graphs that capture relevant macroeconomic and financial information related to the NSE and NASDAQ stock markets and use them to enhance sequential forecasting models.\\
\noindent
    \textbf{(3)} To this end, we propose a time-aware knowledge graph representation technique that leverages a heterogeneous graph convolution operator, along with a temporal point process-based method to model external events. We integrate these embeddings with transformer-based sequential embeddings to obtain accurate rankings of future stock performance over different time horizons.\\    
    \noindent
    \textbf{(4)} We empirically demonstrate that our method outperforms related baselines in ranking stocks across three different holding periods. Additionally, we conduct ablation studies and further experiments to analyze the effectiveness of the different components of our model.

\section{Related work}
\label{section:Related_Work}
Forecasting has traditionally used time-series models that predict future values solely from historical numerical data, assuming that the target variable evolves as a function of its own past. Deep learning techniques using models such as RNNs~\citep{Hewamalage_2021}, LSTMs~\citep{elsworth2020timeseriesforecastingusing, kong2025unlockingpowerlstmlong}, and transformer-based architectures~\citep{zhou2021informerefficienttransformerlong, liu2024itransformerinvertedtransformerseffective} enable the capture of long-range dependencies and complex non-linear patterns that classical models struggle to represent. However, these approaches often assume that all relevant signals are contained within the target series and a limited set of covariates, overlooking the broader context in which the data arise. This limits their ability to handle situations in which external events (e.g., pandemics, geopolitical shifts) or inter-entity interactions (e.g., firm-to-firm relationships) cause abrupt or correlated changes. While recent work has begun to incorporate covariates, there remains a gap in systematically modeling rich, temporally dynamic external influences.

To address this gap, recent studies have incorporated unstructured data sources into forecasting models. These include news articles~\citep{wang2024newsforecastintegratingevent}, social media posts~\citep{karlemstrand2021usingtwitterattributeinformation}, financial reports~\citep{zheng-etal-2019-doc2edag}, and product reviews~\citep{Liu_2023}, which offer valuable contextual signals that complement numerical time series. Text-Guided Time Series Forecasting (TGTSF)\citep{xu2024trendperiodicityguidingtime} uses textual cues, such as channel metadata or news content, to guide predictions. Con \citep{chattopadhyay2025contextmattersleveragingcontextual} combines temporal and textual modalities using cross-attention to achieve improved results.   In a multimodal direction, MM-Forecast~\citep{Li_2024} integrates images and other modalities for temporal-event prediction. 
These approaches typically treat unstructured signals as flat texts, failing to capture the relationships between entities or how they evolve over time. This limits their ability to reason about structured causal chains or model how influences unfold temporally.

Incorporating structural knowledge into learning problems has been studied for natural language problems~\cite{2018:AnnervazChowdhuryDukkipati:LearningBeyodDatasets}. Recently, this technique has been applied to forecasting problems. This includes time-varying categorical variables, continuous external signals, and relational knowledge captured using graphs and knowledge graphs. For example, TimeXer~\citep{wang2024timexerempoweringtransformerstime} enhances transformers with patch-wise and variate-wise attention to handle structured exogenous inputs in multivariate time-series. ChronosX~\citep{arango2025chronosxadaptingpretrainedtime} fuses historical and future auxiliary inputs using dedicated networks to adapt pretrained models without retraining the core models. Beyond tabular features, recent studies have constructed graphs~\citep{cini2023graphdeeplearningtime} and knowledge graphs~\citep{jain2023knowledgegraphrepresentationsenhance} to model prior knowledge and event dynamics that influence targets. This structured context supports better generalization across domains and improves interpretability by linking the forecasts to causally relevant external signals. However, these efforts are often static or limited to coarse-grained relationships and do not explicitly model the evolving temporal interactions between entities. 

Our work addresses this gap by constructing temporal knowledge graphs that encode fine-grained, time-indexed external events and using point-process modeling to dynamically represent their influence on forecasting targets.

\section{Problem formulation}
\label{section:problem_formulation}
Consider a forecasting problem on a multivariate time series 
\begin{displaymath}
\mathcal{P} = \left( \mathcal{P}_i^\tau : i = 1, \dots, N \right)_{\tau},
\end{displaymath}
with $N$ values at each time $\tau$ in some discrete sequence of time points $\tau_1, \tau_2 \dots$. At time $T$, using the history of the series till that time, we wish to forecast
\begin{displaymath}
Y^{T + \Delta} = \displaystyle \mathbf{Y}^\Delta \left( \left( \mathcal{P}_i^\tau \right)_{1 \leq i \leq N}^{\tau_1 < \tau \leq T} \right),
\end{displaymath}
where $Y^{T + \Delta}$ is some property of the $N$ values $ \left( \mathcal{P}_i^{T + \Delta} \right)_{1 \leq i \leq N}$ of the time series that occur at time $T + \Delta$. For the stock forecasting problem, $\mathcal{P}_i^\tau$ denotes the price of stock asset $i$ at time $\tau$, and $Y^{T+\Delta}$ is the ranking of the stock prices of all assets at time $T + \Delta$.

As mentioned earlier, we model the temporal knowledge of external events that can influence this series as a temporal KG. 
For this, we consider the elements $\left( \mathcal{P}_i^\tau \right)_\tau$ of this time series to occur on a set of $N$ nodes of a temporal KG $G$.
While $(\mathcal{P}_i^\tau)$ represents a multivariate series that evolves continuously at every time step, the knowledge graph $G$ has a static component that contains information about all the fixed relations between entities in the graph and a dynamic component that captures the occurrence of events across different time scales. Each event is represented as an edge between the nodes of the graph, valid for a certain time period based on the nature of the event.

To this end, we extended the standard static knowledge graph, consisting of triples of the form $(h, r, t)$, to a temporal knowledge graph by adding a time dimension for the relations, resulting in quintuples of the form $(h, r, t, [\tau_s, \tau_e])$, where $h$ and $t$ are the head and tail entities, respectively, $r$ is the type of relationship between them, and $[\tau_s, \tau_e]$ is the time interval in which the triple remains valid. If the triple is valid for only the time step $\tau$, then it is represented as a quadruple $(h, r, t, [\tau])$.

Thus, the temporal knowledge graph $G$ is a union of subgraphs
\begin{displaymath}
G = G_{\tau_0} \cup \bigcup_{j=1}^{T} G_{\tau_j},
\end{displaymath}
where $G_{\tau_0}$ is the static subgraph and $G_{\tau_j}$ is the set of relations valid at time $\tau_j$.
Therefore, the input $X^T$ to our model at time $T$ is of the form
\begin{displaymath}
X^T = \left( \mathcal{P}_{1 \leq i \leq N}^{T - W < \tau \leq T}, G_{\tau_0}, G_T \right),
\end{displaymath}
where $\mathcal{P}_i^{T - W \leq \tau \leq T}$ is the window of historical stock price data for $W$ previous days of stock $i$ at day $T$, $G_{\tau_0}$ is the non-temporal positive subgraph of $G$, and $G_{\tau_j}$ is the temporal positive subgraph of $G$ at time $\tau_j$.

The objective is to learn embeddings $\mathcal{E}(X^T)$ that can be used to compute scores 
\begin{displaymath}
\widehat{Y}^{T + \Delta} = \mathbf{Y}^{\Delta}_{\mathbf{W}_f}( \mathcal{E}(X^T) )
\end{displaymath}
to rank the best-performing stocks at future time $T + \Delta$. The embedding function $\mathcal{E}$ along with the final prediction layer parameters $\mathbf{W}_f$ are learned using an objective that is a combination of knowledge graph embedding consistency loss and ranking losses that align the output with the ground truth values $\mathcal{P}_i^{T+\Delta}$. The components of this loss function, as well as the detailed equations, are described in Section~\ref{sec:loss_fn} and the Appendix \ref{appdx:sec:loss}.
The learned model is used to predict the top $1$ and $5$ best-performing stocks $\Delta$ time steps in the future. The quality of these predictions is quantified using various relevant metrics, such as return ratios, Sharpe ratios, and accuracy,  as described in Appendix \ref{appdx:sec_metric}.


\begin{figure}[ht]
  \begin{center}
    \centerline{\includegraphics[width=\columnwidth]{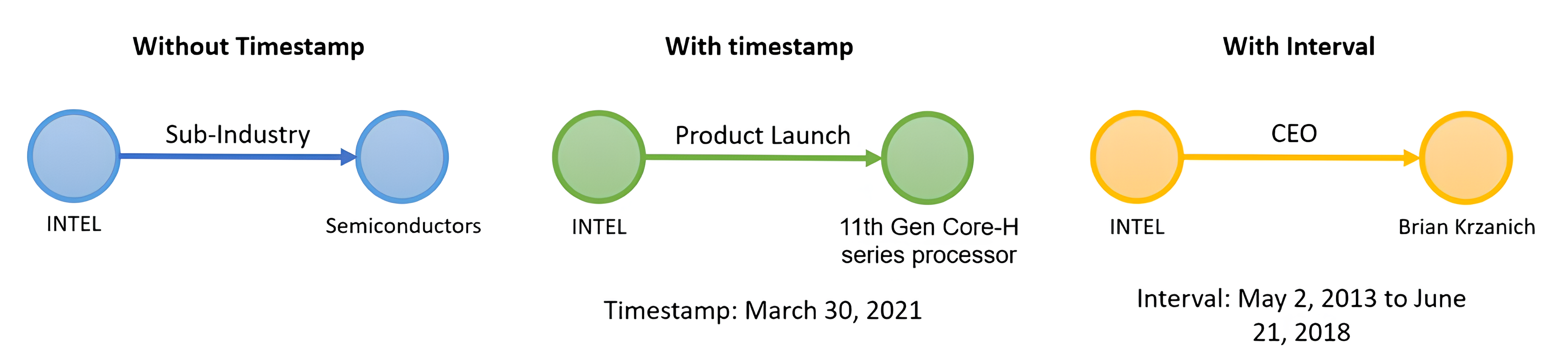}}
    \caption{
      Different types of relations: (left) without a timestamp, (center) with a timestamp, and (right) with a time interval.
    }
    \label{Figure: Relation_Type}
  \end{center}
\end{figure}

\vspace{-8pt}
\section{Temporal Knowledge Graph construction}
The construction of the KGs $\mathsf{STOCKnowledge}$ involved data acquired from various sources pertaining to the NASDAQ and NSE share markets. Temporal KGs are constructed for stocks listed in NASDAQ100, S\&P500, and NIFTY500, spanning a period of 20 years, commencing from January 01, 2003, and concluding on December 30, 2022. 

\textbf{General News} 
News articles were acquired by scraping data from two websites, namely ``Seeking Alpha'' for NASDAQ and ``Money Control'' for NSE, using the Selenium library in Python\footnote{The news articles are obtained by scraping the webpages https://seekingalpha.com/symbol/\{stock\_ticker\}/news for NASDAQ and https://www.moneycontrol.com/company-article/\{stock\_name\}/news/\{stock\_id\}}. A total of 220,645 news articles were collected for the NASDAQ and 272,808 for the NSE. These articles comprise several components, including the headline, body text, publication time, associated stock asset, related sector, and additional taglines, and contain various types of information, such as corporate events, earnings reports, and analyst and investor sentiments. A rule-based pattern-matching system was used to identify events mentioned in the articles using a predefined set of events, which were then added to the knowledge graph.

\begin{figure*}[ht]
  \vskip 0.2in
  \begin{center}
    \centerline{\includegraphics[width=0.95\textwidth]{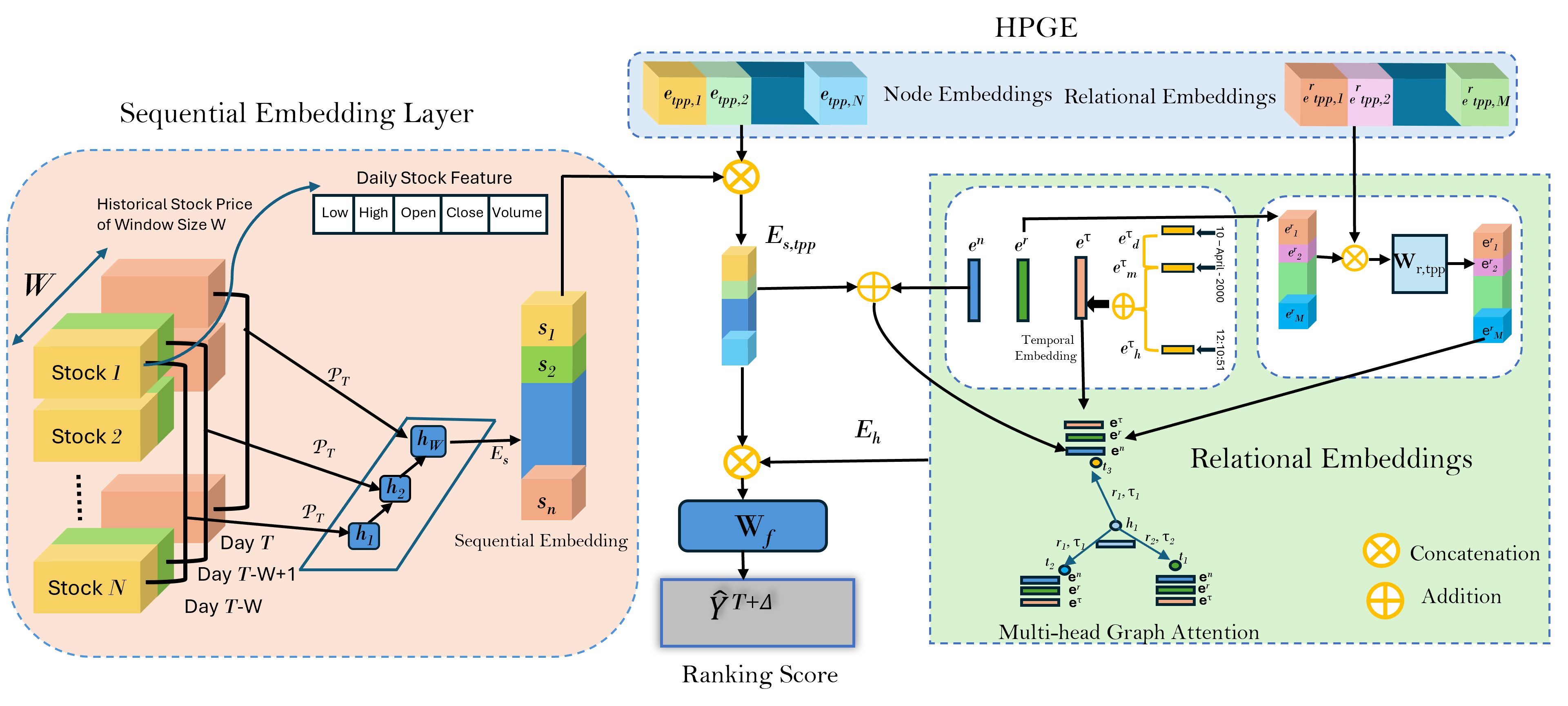}}
    \caption{
      TA-HKGE model architecture
    }
    \label{Figure:TA-HKGE architecture}
  \end{center}
\end{figure*}

\textbf{Financial Statements and Ratios}
The financial statements of companies listed on NASDAQ and NSE were extracted by scraping data from websites ``Macrotrends'' and ``Top Stock Research'' using Selenium\footnote{The financial statements are obtained from www.macrotrends.net and https://www.topstockresearch.com/ for NASDAQ and NSE, respectively.}. This data includes essential components for each stock, including the balance sheet, income statement, cash flow statement, and other financial ratios. Events are derived from financial statements to identify the annual and quarterly trends of these financial parameters. These trends were then incorporated into the knowledge graph as quintuple events, where the interval was defined by the beginning and end of the corresponding financial year or quarter.

\textbf{Macro-Economic Indicators}
Macroeconomic indicators provide a snapshot of the general economic health of a country, including factors such as GDP growth, inflation rate, unemployment rate, trade balance, and fiscal deficit. This information is obtained from ``Macrotrends'' for both the US and India, in which the respective share markets NASDAQ and NSE are located\footnote{The macro-economic indicators are obtained from www.macrotrends.net.}. Trends in indicators were identified as events with intervals defined by the beginning and end of the reported year.

\textbf{Corporate Events}
Corporate events, such as dividend announcements and stock splits, are sourced from Yahoo Finance for the NASDAQ and NSE.

\begin{table}[t]
  \caption{Statistics of the NASDAQ and NSE $\mathsf{STOCKnowledge}$s}
  \label{stat}
  \begin{center}
    \begin{small}
      \begin{sc}
        \begin{tabular}{lcc}
          \toprule
          \toprule
          $\mathsf{STOCKnowledge}$  & NASDAQ & NSE \\
          \midrule
          \midrule
           \# Triples  & 7,568 & 2,736 \\
           \# Quadruples & 52,083 & 44,399 \\
           \# Quintuples & 310,916 & 251,551 \\
           \# Entities & 4,911 & 1,049 \\
           \# Relations & 370,567 & 298,686 \\
           \# Entity types & 12 & 14 \\
           \# Relation types & 56 & 53 \\
           \bottomrule
           \bottomrule
        \end{tabular}
      \end{sc}
    \end{small}
  \end{center}
  \vskip -0.1in
\end{table}

\textbf{Inter-stock Relationship}
Inter-stock relationships can be classified into two types: first-order and second-order relationships. If there is a statement with company $i$ as the subject and company $j$ as the object, then they have a first-order relationship. Furthermore, companies $i$ and $j$ have a second-order relationship if they have statements that share the same object \cite{feng2019temporal}. For instance, ``Apple'' and ``Foxconn'' are in different stages in the production of the product ``iPhone''. This data was extracted from ``Wikibase,'' where the knowledge from ``Wikipedia'' is stored in the form of triples. A preselected set of relations was identified, consisting of $6$ types of first-order relations and $52$ types of second-order relations, which were then filtered from Wikibase.


\textbf{Sector Classification}
Each stock asset is classified into a four-level hierarchical system: Sector, Industry group, Industry, and Sub-Industry by Global Industry Classification Standard (GICS) for NASDAQ, and Macroeconomic Sector, Sector, Industry, and Basic Industry by NSE Indices for NSE. These relations are added as triples with the asset as the head, class as the tail, and classification level as the relation to the constructed knowledge graph.

\section{Proposed Learning Mechanism}
\label{section:model}
The input to the model is $X^T$ at time $T$ consists of
\begin{equation*}
 X^T = \{ \mathcal{P}_i^{\tau \in (T-W, T]}, G_{\tau_0}, G_{T}\}, 
\end{equation*}
where $\mathcal{P}_i^{\tau \in (T-W, T]}$ is the window of historical stock price data for $W$ previous days of stock $i$ at day $T$, $G_{\tau_0}$ is the static subgraph of $\mathsf{STOCKnowledge}$, $G_{\tau_T}$ is the temporal positive subgraph of $\mathsf{STOCKnowledge}$ at time $T$, and $N$ is the total number of stocks in the dataset. 

The objective is to learn an encoder $\mathcal{E}(X^T)$ to obtain output embeddings that can be used to predict the ranking of various stocks at future time steps. We construct this encoder using multiple components: a Sequential Encoder 
\begin{displaymath}
\mathcal{E}_s(\mathcal{P}_i^{\tau \in (T-W, T]}) = E_s 
\end{displaymath}
of the stock prices, a set of temporal process embeddings $e_{\mathrm{tpp}}$ that are learned to optimally model the events as a heterogeneous Hawkes process, and a Relational Encoder
\begin{displaymath}
\mathcal{E}_r(G_{\tau_0}, G_{T}, E_s, e_{\textrm{tpp}}) = E_h.
\end{displaymath}
These are then passed through a prediction layer $\mathbf{Y}^\Delta_{\mathbf{W}_f}$ to obtain a vector of ranking scores for the stocks at time step $T + \Delta$.


\subsection{Sequential Embedding Layer}
Given the strong temporal dynamics of stock markets, it is intuitive to regard a stock's historical performance as the most influential factor in predicting its future trend. Therefore, we first apply a sequential embedding layer to capture the sequential dependencies in the historical price data and technical indicators. Owing to the significant advancements and advantages demonstrated by transformers in processing sequential data, we chose to utilize a transformer model to learn the sequential embeddings. Consequently, we feed the historical time series data $\mathcal{P}_i^T$ of stock $i$ at time step $T$ into the Transformer model, which generates the sequential embedding $s_i^T$ for each stock as
$s_i^T = \mathsf{Transformer} \left( \mathcal{P}_i^T \right).$
The Transformer model processes the entire historical sequence simultaneously, effectively capturing long-range dependencies. 
We employed the encoder-only architecture of the Transformer~\citep{vaswani2017attention} to obtain sequential embeddings.

\subsection{Temporal Process Embeddings}
We then consider relations on the dynamic knowledge graph as events of a heterogeneous Hawkes process on the graph and learn node and relation embeddings using the Heterogeneous Hawkes Process for Dynamic Heterogeneous Graph Embedding (HPGE) algorithm~\citep{ji2021dynamicheterogenousgraphhawkes}. For event $(h, r, t, \tau)$, we consider the intensity function defined as
\begin{align*}
    & \tilde{\lambda}(h, r, t, \tau) = \mu_r(h, t) + \gamma_1 \sum_{r', t', \tau' \in \mathcal{N}_{<\tau}(h)} m_1(h, r', t', \tau')  
    \\
    & \qquad \qquad + \gamma_2 \sum_{r', h', \tau' \in \mathcal{N}_{<\tau}(t)} m_2(t, r', h', \tau'),
\end{align*}
where $\mu_r$ is the relation type $r$-specific base intensity function between $h$ and $t$, and $m_1$ and $m_2$ are the mutual intensities between the entities, summing over all relation types $r'$ in the past neighborhoods of the nodes $h$ and $t$ with respective neighbors $t'$ and $h'$.

Node embeddings $e_{\text{tpp}, i}$ and relation embeddings $e^r_{\textrm{tpp}, j}$ are initialized and the base intensity $\mu_r(h, t)$ is taken as 
\begin{equation*}
     \left\Vert \sigma \left(\mathbf{W}_e \cdot e_{\text{tpp}, h}\right) + e^r_{\text{tpp}, r} - \sigma \left(\mathbf{W}_e \cdot e_{\text{tpp}, t} \right)  \right\Vert^2,
\end{equation*}
where $\mathbf{W}_e$ is a dense layer, and the activation function $\sigma$ is LeakyReLU. The embeddings and weights are optimized by maximizing the intensity function for the positive samples and minimizing the intensity function for the negative samples. The loss function and mutual information terms are defined for the HPGE algorithm~\cite{ji2021dynamicheterogenousgraphhawkes}.
The node embeddings obtained from the sequential layer and temporal process model are concatenated to create new node embeddings $E_{s, \text{tpp}} = \left( s_i \| e_{\text{tpp}, i} \right)_i$.

\subsection{Relational Embeddings}
To obtain the relational embedding, each entity $i$ in the knowledge graph is assigned a learnable embedding $e^n_i$ randomly initialized from $\mathcal{N}(0, 1)$. The node embeddings $E_{s, \text{tpp}}$ are added to the embeddings of the nodes corresponding to the selected stock assets. Together, this creates an entity feature matrix $\mathbf{e}^n = \left( e^n_i \right)_{i = 1, \dots, N}$.
Similarly, features for the set of relation types are defined as embeddings $\left( e^r_j \right)_{j = 1, \dots, M}$, where $M$ is the number of relation types, and are initialized from $\mathcal{N}(0,1)$. Each of these is concatenated with the corresponding relation embedding $e^r_{\text{tpp}, j}$ obtained from the temporal process embedding method and projected to its original dimension to obtain the relational embeddings
\begin{displaymath}
\mathbf{e}^r = \mathbf{W}_{r, \text{tpp}} \cdot \left[ e_r \Vert e^r_{\text{tpp}} \right].
\end{displaymath}
Additionally, embeddings are defined for the month $e^\tau_m$, day $e^\tau_d$ and hour $e^\tau_h$, with weights initialized from $\mathcal{N}(0,1)$. These are then summed to obtain the temporal embedding for the start timestamp $\tau_s$, denoted by $\mathbf{e}^\tau$.
These node embeddings $\mathbf{e}^n$, along with edge attributes $\mathbf{e}^e = \mathbf{e}^r + \mathbf{e}^\tau$ are passed through two layers of Heterogeneous Edge-featured Graph Attention (HEAT)~\citep{mo2021heterogeneous} and two linear layers to obtain the final combined relational embedding $E_h$ for all stocks.

\subsection{Prediction Layer}
The obtained embeddings are passed through a prediction layer to obtain a vector of ranking scores
\begin{equation}
    \hat{Y}^{T+\Delta} = \frac{\exp(\mathbf{W}_f(E_{s, \text{tpp}} \| E_h )+b_f)}{\sum_{n=1}^{N} \exp(\mathbf{W}_f(E_{s, \text{tpp}} \| E_h )+b_f)}
\end{equation}
that acts as the predicted rank of the given stocks in the non-increasing order of their potential return on investment from time $T$ to $T + \Delta$.

\setlength{\tabcolsep}{5.2pt} 
\begin{table*}[]
    \centering
    \caption{Top 5 Performance comparison of TA-HKGE against baseline models, reported as averages over all phases of the corresponding datasets NASDAQ100, S\&P500, and NIFTY500. Best results for each holding period are shown in \textbf{bold}.}
    \label{TOp5_Results}
    \begin{scriptsize}
    \begin{tabular}{l p{1.3cm} ccccc | ccccc | ccccc}
        \hline
        \hline
        \multirow{2}{*}{$\Delta$} & \multirow{2}{*}{Model} 
        & \multicolumn{5}{c|}{NASDAQ100} 
        & \multicolumn{5}{c|}{S\&P100} & \multicolumn{5}{c}{NIFTY500} \\ \cline{3-17}
         &  
         & IRR & AIRR & SR & NDCG & ACC & IRR & AIRR & SR & NDCG & ACC & IRR & AIRR & SR & NDCG & ACC \\ 
        \hline
        \hline
        \multirow{6}{*}{1} 
        & TRANSF 
        & 0.080 & 22.498 & 0.090 & 0.315 & 5.556 &  0.065 & 17.900 & 0.030 & 0.208 & 1.103 & 0.098 & 28.173 & -1.020 & 0.222 & 1.372 \\
        & STHGCN 
        & 0.082 & 22.971 & 0.117 & 0.309 & 4.931 & 0.052 & 13.969 & -0.020 & 0.204 & 0.800 &  0.069 & 19.114 & 0.060 & \textbf{0.310} & \textbf{4.740}\\
        & STHAN 
`       & 0.069 & 19.114 & 0.057 & 0.308 & 4.744 & 0.059 & 16.013 & 0.007 & 0.205 & 0.914 & 0.112 & 32.582 & -0.720 & 0.220 & 1.433 \\
        & GCNKG 
        & 0.100 & 28.741 & \textbf{0.135} & 0.312 & 5.136 & 0.070 & 19.240 & 0.020 & 0.207 & 1.072 &  0.109 & 31.538 & -1.080 & 0.218 & 1.233 \\
        & CI-STHPAN
        & 0.105 & 30.120 & 0.118 & 0.355 & 9.420 &  0.075 & 20.410 & 0.038 & 0.218 & 1.680 & 0.140 & 42.600 & -0.420 & 0.232 & 2.980 \\
        \rowcolor[HTML]{FFDAB9}
        & \textbf{TA-HKGE} 
        & \textbf{0.110} & \textbf{31.920} & 0.100 & \textbf{0.397} & \textbf{13.150} & \textbf{0.080} & \textbf{22.320} & \textbf{0.059} & \textbf{0.228} & \textbf{2.300} & \textbf{0.170} & \textbf{53.420} & \textbf{0.050} & 0.246 & 4.630 \\
        
        \hline
        \hline

        \multirow{5}{*}{5} & TRANSF
        & 0.422 & 23.675 & 0.551 & 0.316 & 5.736 & 0.276 & 14.882 & 0.420 & 0.206 & 0.942 & 0.501 & 28.633 & 0.071 & 0.220 & 1.317 \\
        & STHGCN
        & 0.484 & 27.559 & \textbf{0.623} & 0.318 & 6.114 & 0.328 & 17.961 & 0.482 & 0.208 & 1.242 & 0.502 & 28.737 & 0.049 & 0.220 & 1.411\\
        & STHAN
        & 0.489 & 27.917 & 0.611 & 0.318 & 6.197 & \textbf{0.363} & \textbf{20.028} & \textbf{0.517} & 0.209 & 1.217 & 0.489 & 27.867 & 0.031 & 0.217 & 1.322 \\
        & GCNKG
        & 0.417 & 23.333 & 0.545 & 0.316 & 5.744 & 0.308 & 16.801 & 0.480 & 0.206 & 1.025 & 0.516 & 29.757 & 0.065 & 0.219 & 1.356 \\
        & CI-STHPAN
        & 0.520 & 30.180 & 0.570 & 0.355 & 9.240 & 0.240 & 12.900 & 0.330 & 0.214 & 1.860 & 0.585 & 34.120 & 0.320 & 0.223 & 1.740 \\
        \rowcolor[HTML]{FFDAB9}
        & \textbf{TA-HKGE}
        & \textbf{0.620} & \textbf{36.540} & 0.597 & \textbf{0.397} & \textbf{13.330} & 0.180 & 9.480 & 0.200 & \textbf{0.222} & \textbf{2.900} & \textbf{0.650} & \textbf{38.610} & \textbf{0.652} & \textbf{0.227} & \textbf{2.130} \\

        \hline
        \hline

        \multirow{5}{*}{20} & TRANSF
        & 1.711 & 23.829 & 1.382 & 0.315 & 5.950 & 1.126 & 15.154 & 1.142 & 0.205 & 1.239 & 2.310 & 33.351 & 1.225 & 0.221 & 1.442 \\
        & STHGCN
        & 1.718 & 23.935 & 1.371 & 0.314 & 6.200 & 1.490 & 20.460 & \textbf{1.340} & 0.209 & 1.619 & 2.358 & 34.141 & 1.280 & 0.221 & 1.586 \\
        & STHAN
        & \textbf{2.013} & 28.340 & \textbf{1.540} & 0.316 & 6.819 & 1.376 & 18.846 & 1.165 & 0.211 & 1.567 & 2.388 & 34.657 & 1.330 & 0.220 & 1.592 \\
        & GCNKG
        & 1.404 & 19.654 & 1.126 & 0.319 & 6.372 & 1.168 & 15.794 & 1.085 & 0.206 & 1.253 & \textbf{2.470} & \textbf{36.06} & 1.294 & 0.220 & 1.575\\
        & CI-STHPAN
        & 1.720 & 24.180 & 0.920 & 0.355 & 9.640 & 1.360 & 18.740 & 0.980 & 0.220 & 2.420 & 2.020 & 28.900 & 1.560 & 0.228 & 2.050 \\
        \rowcolor[HTML]{FFDAB9}
        & \textbf{TA-HKGE}
        & 2.000 & \textbf{28.548} & 0.718 & \textbf{0.391} & \textbf{12.990} & \textbf{1.560} & \textbf{21.530} & 0.907 & \textbf{0.235} & \textbf{3.530} & 1.550 & 21.380 & \textbf{1.880} & \textbf{0.236} & \textbf{2.670}\\

        \hline  
        \hline
    \end{tabular}
    \end{scriptsize}
\end{table*}
\subsection{Loss Function}
\label{sec:loss_fn}
The loss function is defined as the sum of the following four terms:
\begin{align}
    \mathcal{L} &= \alpha_1 \mathcal{L}_1(h, r, t, \tau) + \alpha_2 \mathcal{L}_2(\{\hat{Y}\}, \{Y\}) \nonumber \\
    & \qquad + \alpha_3 \mathcal{L}_3(\{\hat{Y}\}, \{D\}) + \alpha_4 \mathcal{L}_4(\{\hat{Y}\}, \{R\}).
\end{align}
Here $\mathcal{L}_1(h, r, t, \tau)$ is the Knowledge Graph Embedding loss, which ensures consistency of the learned node and temporal embeddings by minimizing
\begin{equation*}
    \mathcal{L}_1(h, r, t, \tau) = \| e^n_h + e^r + e^\tau - e^n_t \|,
\end{equation*}
where $\mathcal{L}_2$ is the listwise loss function ApproxNDCG~\citep{bruch2019revisiting} comparing the actual and predicted ranks, $\mathcal{L}_3$ is a binary cross-entropy loss between the prediction and an indicator variable whether the stock price moved in the upward direction, and finally, $\mathcal{L}_4$ is another cross-entropy loss between the predicted ranking score and an indicator whether the stock is in the top 5 with respect to the IRR metric. The detailed expressions for each of the above losses are provided in Appendix \ref{appdx:sec:loss}.

\section{Experimental setup}
\label{section:Experiemental_Setup}

\begin{table}[t]
  \caption{Statistics of the three datasets - NASDAQ100, S\&P500 and NIFTY500.}
  \label{dataset_stats}
  \begin{center}
    \begin{small}
      \begin{sc}
        \begin{tabular}{lccccr}
          \toprule
          \toprule
            - & NASDAQ100 & S\&P500 & NIFTY500 \\
            \midrule
            \midrule
            \#Assets & 83 & 442 & 336 \\
            \#Days & 2800 & 2800 & 2800 \\
           \bottomrule
           \bottomrule
        \end{tabular}
      \end{sc}
    \end{small}
  \end{center}
  \vskip -0.1in
\end{table}

\subsection{Datasets} 
Three datasets are created, containing a subset of stocks from NASDAQ and NIFTY, with datasets NASDAQ100 and S\&P500 from the NASDAQ market and NIFTY500 from NSE, using Yahoo Finance. Each dataset consists of historical price movements and a constructed knowledge graph valid for each trading day. The historical price movements include features Close, Open, Low and High prices, and trading volume for each trading day from 30 January 2012 to 20 October 2022. The statistics are presented in Table~\ref{dataset_stats}.

The model is given as input a context window of stock data $X_T$ at time $T$, which consists of the historical stock price $\mathcal{P}^{\tau \in (T - W, T]}$ and the positive subgraph $G_T$ from the knowledge graph at time point $T$. To avoid data leakage and minimize distribution drift, the values in the historical price window are normalized by dividing them by the last value in the previous window.

To learn embeddings using the heterogeneous Hawkes process, we transformed the data tuples into the format (head, head\_type, tail, tail\_type, relation, timestamp), with a separate tuple generated for each month within the range of validity of the relation.

\subsection{Training Setup} 
The value of $\Delta$, the number of days in the future to be considered for prediction, is set as 1, 5, or 20, corresponding to daily, weekly, or monthly holding periods, respectively. We divided the dataset into 24 smaller overlapping datasets to train and evaluate the model separately in different phases by shifting a 600-day sliding window over the entire trading period, starting from a 400-day window. Each phase is divided into a training period of 450 trading days, a validation period of 50 days, and a testing period of 100 days, with the first few phases starting with a 250-day training period. The calculated metrics were obtained by averaging the results from all test sets across the 24 phases. The metrics reported are described in the Appendix \ref{appdx:sec_metric}.

\begin{figure*}[ht]
  \centering
  \includegraphics[width=1\linewidth, keepaspectratio]{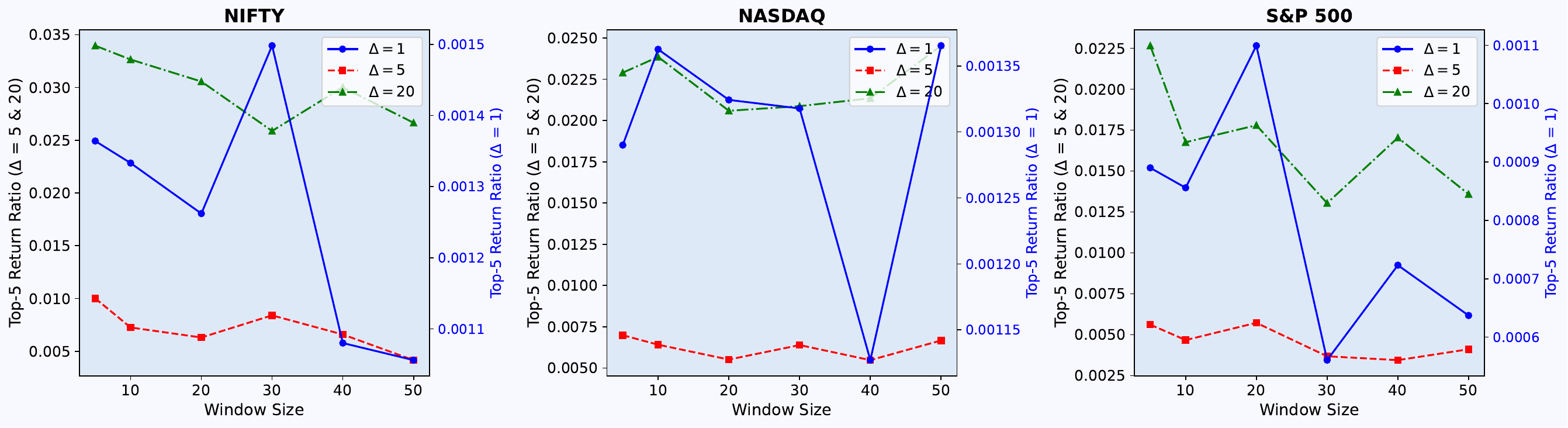}
  \caption{
    Effect of sliding window size $W$ on Top-5 return ratio across different holding periods $\Delta$.
  }
  \label{fig:rr_vs_window}
\end{figure*}

\begin{figure*}[ht]
  \centering
  \includegraphics[width=1\linewidth, keepaspectratio]{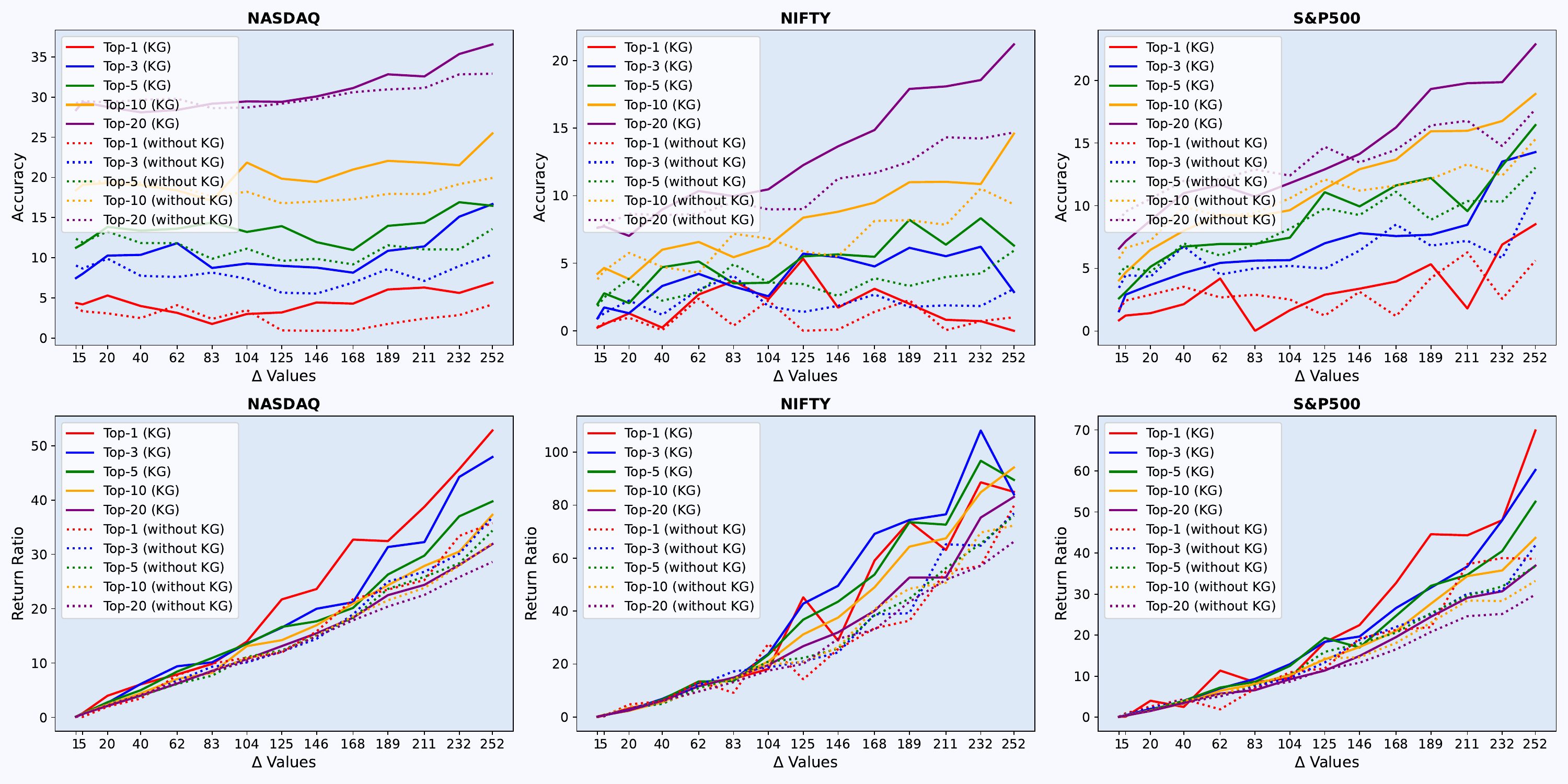}
  \caption{
    Analysis of accuracy and return ratio across varying holding periods $\Delta$ for each dataset, comparing models with and without knowledge graph integration}
  \label{fig:kg_withoutkg}
\end{figure*}
\subsection{Baselines}
First, we consider a simple transformer (TRANSF)~\citep{vaswani2017attention} acting on historical price information $\mathcal{P}$. Next, we consider the Spatio-Temporal Hypergraph Convolution Network (STHGCN)~\citep{sawhney2020spatiotemporal}, which uses a gated temporal convolution over hypergraphs constructed from stock asset sector classifications, and the Spatio-Temporal Hypergraph Attention Network (STHAN-SR)~\citep{sawhney2021stock}, which models stock dependencies via hypergraphs to capture higher-order relations, with a hyperedge indicating the presence of a first- or second-order relation. Channel-independent spatiotemporal hypergraph pretrained Attention Networks(CI-STHPAN)~\citep{xia2024ci} is a two-stage stock selection model that pretrains a transformer on stock time-series data and incorporates a hypergraph attention network to model inter-stock relations. It constructs channel-independent dynamic hypergraphs using DTW-based similarity and is fine-tuned with a ranking-based objective for stock selection. Finally, we consider GCNKG, a two-layer GCN that considers a homogeneous directed graph version of the current time snapshot of the constructed knowledge graphs to obtain node embeddings. These are then concatenated with the sequential embeddings and used to calculate the prediction scores. The loss function used for training the baselines is a linear combination of the pair-wise ranking loss (6) and direction loss (9) defined in the Appendix \ref{appdx:sec:loss}. We do not include LLMs as baselines because they cannot effectively model large temporal knowledge graphs owing to context length limitations and the cost associated with long-context inputs. Moreover, they lack explicit mechanisms for structured, time-aware relational reasoning required for our task.
\begin{figure*}[ht]
\begin{center}
\centerline{\includegraphics[width=1\linewidth,keepaspectratio]{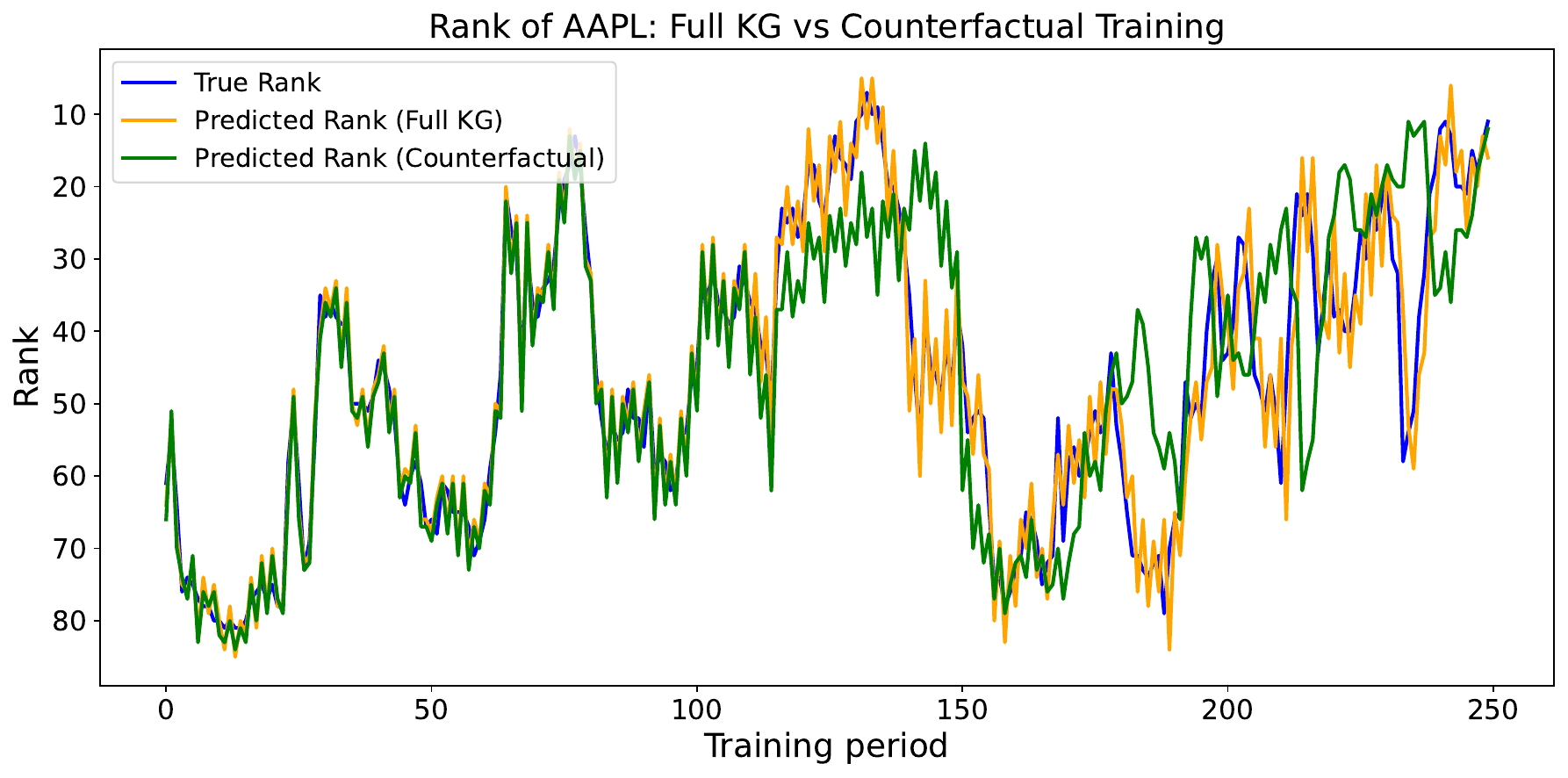}}
\caption{Rank trajectory of AAPL under full KG and counterfactual KG training, shows gradual degradation after relation removal.}
\label{Rank_counterfactual}
\end{center}
\end{figure*}

\subsection{Overall Performance} 
\label{sec:overall_performance}
In Tables \ref{TOp5_Results} and Table \ref{Top1_Results} (Appendix), we compare TA-HKGE with five baseline methods across all holding periods ($\Delta\in\{1,5,20\}$). The proposed \textbf{TA-HKGE} consistently achieves the strongest or near-strongest ranking quality (NDCG/ACC) and investment returns (IRR/AIRR) across both the NASDAQ and NSE markets. Purely sequential models (e.g., \textsc{TRANSF}) perform well when market dynamics are driven by historical momentum; however, their performance drops when external events cause changes in market behavior.
Graph-based baselines with fixed relational structures (e.g., \textsc{STHGCN}/\textsc{STHAN}/\textsc{GCNKG}) capture cross-asset dependencies to some extent, yet are constrained by static or weakly time-aware representations.
In contrast, TA-HKGE improves robustness by jointly modeling temporal external events and heterogeneous temporal knowledge graph structure alongside price dynamics, leading to stronger and more stable gains across short , medium , and long period forecasts.

\section{Ablations and Additional Experiments}
\label{sec:ablations}
\subsection{Impact of Sliding window} 
Our method integrates historical stock prices with temporal events and relationships to forecast future trends. To analyze the effect of the historical context length, we varied the sliding window size $W$ used in the sequential embedding layer. Moderate window sizes generally yield stronger results, whereas very large values of $W$ lead to a gradual decline in the return ratio. This behavior suggests that excessively long price histories may introduce outdated or noisy information that dilutes relevant temporal signals and negatively impacts performance.

\subsection{Impact of Holding Period ($\Delta$) and Knowledge Graph Integration} Figure~\ref{fig:kg_withoutkg} shows that return ratios improve with increasing holding periods, indicating stable performance across different horizons. In addition, models with knowledge graph integration consistently outperform those without KG in both return ratio and accuracy across all datasets. The growing performance gap for larger $\Delta$ further demonstrates the importance of external relational information in robust long-term forecasting.

We wish to study the contribution of the individual components of our method TA-HKGE. We compared multiple variants of TA-HKGE. For the first WOTPP, we removed the embeddings obtained from temporal process modeling and only considered sequential and relational embeddings. For the second variant, LSTM*, we replace the transformer based sequential embedding in the WOTPP with an LSTM. Next, we consider WOSEQ, in which the sequential embedding update for the stock assets as input for the relational embedding layer is removed, along with the embeddings obtained from the temporal point process model. Finally WOHK, the temporal point process embeddings and heterogeneous transformation of entity features are removed . Table \ref{change} in the Appendix provides detailed results, shows that each component of our architecture contributes meaningfully to overall performance of the algorithm.

We further evaluated the robustness of TA-HKGE under longer training, validation, and testing horizons. where market drifts in stock prices over time and evolving external events pose additional challenges. To measure the extent of this effect, we conducted experiments with different larger training, validation, and testing periods for each phase. The results are presented in Table\ref{days_increase} in the Appendix, shows the model performance across different training periods. However, the model consistently outperform risk-free returns, as reflected by the Sharpe ratio.

We include an additional analysis on metric variance across phases and the effect of trading the stock at best and worst case times during the day in the Appendix \ref{appdx:sec:exps} Table \ref{ablation1}.

\section{Counterfactual Interpretability: How External Relations Influence Return Ratios}
To understand how and why external knowledge graph relations influence TA-HKGE’s return predictions, a key advantage of TA-HKGE is that it is auditable: predictions are not only accurate but can be traced back to time-stamped typed knowledge graph (KG) relations. Therefore, we evaluate interpretability via a controlled counterfactual intervention that removes specific KG relation types while keeping the historical price trajectories fixed and measures the change in return ratio (RR) along with relation-level attention.

We train (i) with the original full temporal KG and (ii) with a counterfactual KG, where selected relations are removed. Thus, any performance drop can be attributed directly to missing external evidence.

\subsection{NASDAQ Case Study: Apple Inc. (AAPL)}
In the NASDAQ \textsf{STOCKnowledge} graph, the Apple Inc. (AAPL) connects to heterogeneous, time-aware relationships that capture both structural and event-driven external information.
Under the original full KG, TA-HKGE consistently ranks AAPL among the Top-1 and Top-5 stocks and achieves stable investment return ratios (IRR) across daily, weekly and monthly holding periods.

To assess which external relations drive the gains, we removed two complementary relation types during training. 

(i) DECLARES\_DIVIDEND: a discrete, high-impact corporate event with a well-defined timestamp; A short-horizon event and receives high attention around event windows.

(ii) INCREASE\_YoY: a longer-horizon financial indicator that encodes persistent performance trends and can influence medium/long-term valuation expectations.

Following this removal, IRR decreases for short horizons ($\Delta = 1, 5$). For longer horizons ($\Delta = 20$), IRR shows a gradual decline, dropping from \textbf{$0.73$} (Full KG) to \textbf{$0.67$} (counterfactual KG).
Figure~\ref{Rank_counterfactual} shows the corresponding rank trajectories: before intervention, both models behave similarly, whereas after relation removal, the counterfactual model progressively diverges.

\begin{figure*}[ht]
\begin{center}
\centerline{\includegraphics[width=1\linewidth,keepaspectratio]{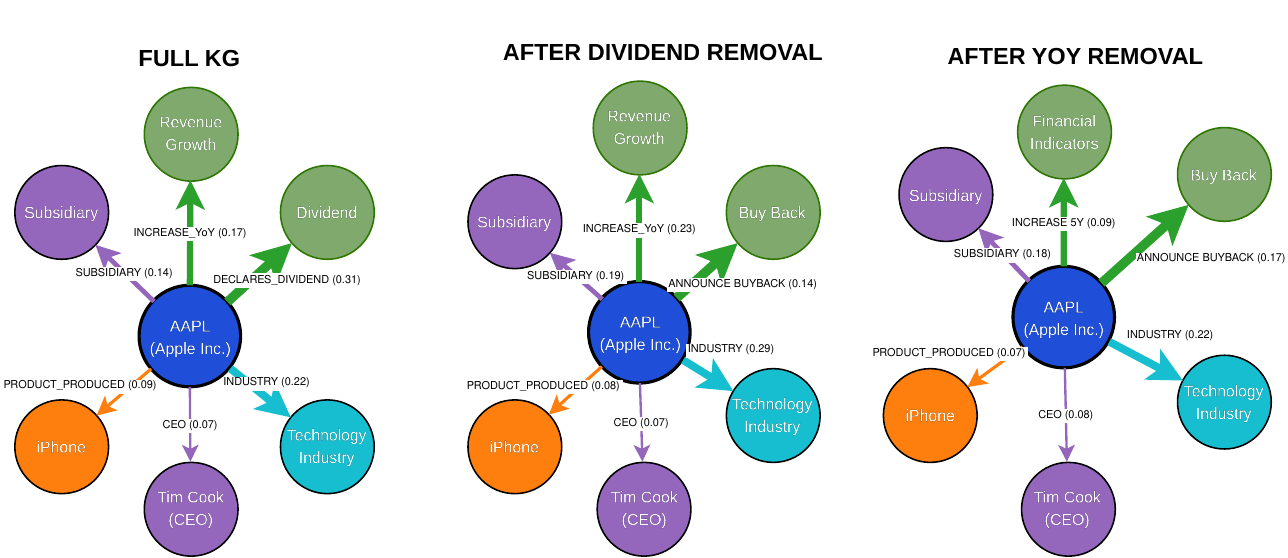}}

\caption{Top attended relations for AAPL over time before and after counterfactual intervention.}
\label{counterfactual_attention}
\end{center}
\end{figure*}
\subsection{Attention-Based Explanation} Figure~\ref{counterfactual_attention} analyzes the relation-level attention learned by the heterogeneous graph attention layers.
Under the full KG, attention focuses on dividend-related corporate actions and financial growth indicators near event timestamps.
After removing "DECLARES\_DIVIDEND" and "INCREASE\_YoY", attention shifts toward the remaining structural relations, which are less informative.

This attention shift aligns with the observed degradation in return ratios and provides a mechanistic explanation for the drop in RR: TA-HKGE leverages the KG to localize which external events matter at specific times, rather than treating contextual information as unstructured noise.

\subsection{Discussion}
This case study indicates that TA-HKGE’s performance are not solely due to increased model capacity but also to the meaningful integration of heterogeneous, time-aware external knowledge.
Different relation types contribute at different temporal scales, and counterfactual analysis provides model-consistent evidence of when and which external relations influence the model's decision.

The detailed analysis of the model’s time complexity is in Appendix \ref{appdx:timecomplexity}.

\section{Conclusion}
\label{conclusion}
This study, for the first time, addresses forecasting problems by incorporating dynamic external influences, a departure from the predominant focus on historical data in existing techniques. We constructed temporal KGs called $\mathsf{STOCKnowledge}$ that capture the temporal and heterogeneous nature of the event data.
Our main contributions are the development of learning mechanisms using heterogeneous graph attention networks, temporal process models, and time-aware translation-based knowledge graph embedding techniques to address forecasting with external influence. Although we demonstrated our results on the stock market data, the proposed learning algorithms are
sufficiently versatile to be applied to forecasting problems across various fields. 

The dataset and relevant codes for the experiments can be accessed at 
\url{https://sml.csa.iisc.ac.in/Blog/stocks/index.html}



\bibliography{example_paper}
\bibliographystyle{icml2026}

\newpage
\appendix
\onecolumn

\begin{center}
    \text{\Large \textbf{Predictive AI with External Knowledge Infusion:}}\\ \text{\Large \textbf{Datasets and Benchmarks for Stock Markets}} \\
    \text{\Large \textbf{\textit{Appendix}}}
\end{center}

\section{Loss functions}
\label{appdx:sec:loss}
\subsection{KGE loss function}
Time-aware knowledge graph embeddings is learned using a translation-based model-type loss function, in which the element-wise addition of the head entity, relation embedding, and temporal embedding should be close to the tail embedding.
\begin{align}
    \mathcal{L}_1(h, r, t, \tau) = \| e^n_h + e^r + e^\tau - e^n_t \|.
\end{align}

\subsection{Listwise ranking}
Earlier works \cite{feng2019temporal, sawhney2020spatiotemporal, sawhney2021stock} employed the following pairwise loss function as the objective.
\begin{equation}
    \sum_{i=0}^N \sum_{j=0}^N \max \left\{ 0, -\left( \hat{Y}_i^{T+\Delta} - \hat{Y}_j^{T+\Delta} \right) \left( Y_i^{T+\Delta} - Y_j^{T+\Delta} \right) \right\},
    \label{pairwiseloss}
\end{equation}
where $\hat{Y}_i^{T+\Delta}$ is the predicted ranking score, and $Y_i^{T+\Delta}$ is the actual ranking score for stock $i$. This approach computes the loss for each pair of scores separately, which might be too loose as an approximation of the NDCG and MAP performance measures. In contrast, the listwise loss function ApproxNDCG \cite{bruch2019revisiting} used in the listwise approach can better represent the performance measures. The listwise ranking loss function is given by,

\begin{align}
    \mathcal{L}_2(\{\hat{Y}\}, \{Y\}) &= - \frac{1}{\text{DCG}(\hat{y}, \hat{y})} \sum_{i} \frac{2^{\hat{y}_i} - 1}{\log_2(1 + \text{rank}_i)}, \\
\text{with ~rank}_i &= 1 + \sum_{j \neq i} \frac{1}{1 + \exp\left(\frac{-(y_j - y_i)}{\text{temperature} }\right)},
\end{align}
where $y$ is the predicted ranking score and $s$ is the actual score, which is the IRR for each stock from day $T$ to $T + \Delta$.

\subsection{Cross-Entropy losses}
Let $D^{T+\Delta}_i$ be the binary movement direction from the closing price of stock $i$ from day $T$ to $T+\Delta$, in which 1 denotes rise, and 0 denotes fall, and it is expressed as $ D^{T+\Delta}_i = 1(p^i_{T+\Delta} \geq p^i_{T}) $, where $p^i_\tau$ is the closing price of stock $i$ on the day $\tau$. The movement ranking loss is defined as the binary cross-entropy between $\hat{Y}^{T+\Delta}_i$ and $D^{T+\Delta}_i$.
\begin{align}
     \mathcal{L}_3(\{\hat{Y}\}, \{D\}) = -[D \log(\hat{Y}) + (1-D) \log(1-\hat{Y})]
     \label{dirloss}
\end{align}
In addition, another cross-entropy loss is calculated between the predicted ranking score and stocks that appeared in the top $K$ ranks. Such that $ R^{T+\Delta}_i = 1(IRR_i \text{is in Top 5}) $ is the truth label.
\begin{align}
     \mathcal{L}_4(\{\hat{Y}\}, \{R\}) = -[R \log(\hat{Y}) + (1-R) \log(1-\hat{Y})]
\end{align}

\begin{figure*}
\centerline{\includegraphics[width=\linewidth]{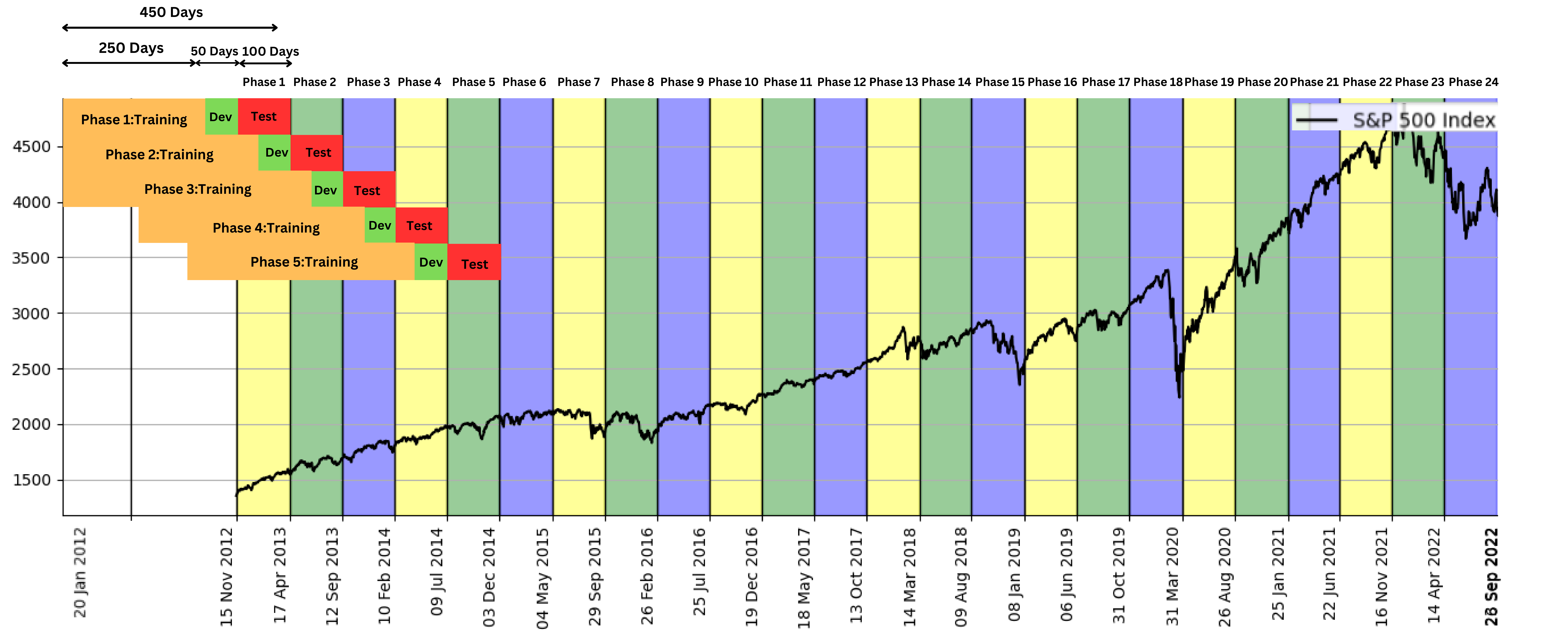}}
\caption{[Best viewed in color] The experimental setup for model training is illustrated across 24 phases. In each phase, the model undergoes a training period of 250 days, followed by a validation period of 50 days and, subsequently, a testing period of 100 days.}
\label{Figure: Phase_training}
\end{figure*}

\section{Evaluation metrics}
\label{appdx:sec_metric}
As illustrated in Figure~\ref{Figure: Phase_training}, we divided the dataset into 24 smaller overlapping datasets to train and evaluate the model separately in different phases by shifting a 600-day sliding window over the entire trading period, starting from a 400-day window. Each phase is divided into a training period of 450 trading days, a validation period of 50 days, and a testing period of 100 days, with the first few phases starting from a 250-day training period. The calculated metrics are obtained by averaging the results from all the test sets across the 24 phases.

\label{appdx:sec:metrics}
\subsection{Investment Return Ratio (IRR)}
The Cumulative Investment Return Ratio (IRR) is defined as
\begin{equation*}
    \text{IRR} = \frac{\text{Final Value} - \text{Initial Value}}{\text{Initial Value}} \times 100,
\end{equation*}
The percentage difference between the value of the investment at the start and end of the holding period.

\subsection{Annualized Investment Return Ratio (AIRR)}
The Annualized Investment Return Ratio (AIRR) is defined as
\begin{equation}
    \text{AIRR} = (1 + \text{ROI})^{\frac{252}{\Delta}} - 1
    \label{irr},
\end{equation}
where $252$ is the average number of trading days in a year, $\Delta$ is the number of holding days, and ROI is the Return On Investment defined as IRR$/100$. AIRR quantifies the average annual rate of return achieved on an investment, considering the compounding effect over time.

\subsection{Sharpe Ratio (SR)}
The Sharpe ratio measures the risk-adjusted return of an investment and can be calculated using the following formula:
\begin{align}
    \text{SR} &= \frac{R_p - R_f}{\sigma_p}
\end{align}
where $R_p$ represents the average return of the investment, $R_f$ the risk-free rate of return, and $\sigma_p$ the standard deviation of the investment's returns. The Sharpe ratio quantifies the excess return earned per unit of risk. It is commonly used to assess the attractiveness of investment opportunities based on risk-adjusted performance. 
The risk-free rates are calculated for each phase by averaging the 6-month treasury-bill yield obtained from ``www.treasurydirect.gov" for the USA and ``www.investing.com" for India.

\subsection{Normalized Discounted Cumulative Gain (NDCG)}
Discounted Cumulative Gain (DCG) is a metric used to evaluate the quality of a ranked list. It can be calculated as
\begin{align}
    \text{DCG@}k &= \sum_{i=1}^k \frac{\text{rel}_i}{\log_2 (i+1)},
\end{align}
where $\text{rel}_i$ represents the relevance score of the item at position $i$ in the ranked list. The DCG assigns higher weights to more relevant items at the top of the list by dividing the relevance score by the logarithm of their position. The DCG accumulates these weighted scores up to position $k$ to measure the overall quality of a ranked list. The Ideal Discounted Cumulative Gain (IDCG) is a reference value used to evaluate the quality of a ranked list, where $\text{rel}_i$ represents the relevance score of the item at position $i$ in the ideal ranking order. 
The Normalized Discounted Cumulative Gain (NDCG) is a metric used to evaluate the quality of ranked lists. It can be calculated using the following formula:
\begin{align}
    \text{NDCG@}k &= \frac{\text{DCG@k}}{\text{IDCG@k}}
\end{align}
NDCG provides a normalized measure of ranking performance, ranging from 0 to 1, with higher values indicating better ranking quality.

\subsection{Accuracy (ACC)}
Accuracy represents the ratio of correct predictions to the total number of predictions made by the model. It provides an overall measure of the model's ability to correctly identify the top $k$ stocks at each time point. 

\begin{align}
    \text{ACC@}k &= \frac{|\text{Top k Predictions} \cap \text{Top k Stocks}|}{|\text{Top k Stocks}|} \times 100
\end{align}

This Table \ref{Top1_Results} corresponds to the main results discussed in the Overall Performance \ref{sec:overall_performance} of this paper.

\setlength{\tabcolsep}{5.2pt} 
\begin{table}[ht]
    \centering
    \caption{Top 1 Performance comparison of TA-HKGE against baseline models, reported as averages over all phases of the corresponding datasets NASDAQ100, S\&P500, and NIFTY500. Best results for each holding period are shown in \textbf{bold}.}
    \label{Top1_Results}
    \begin{scriptsize}
    \begin{tabular}{l p{1.3cm} ccccc | ccccc | ccccc}
        \hline
        \hline
        \multirow{2}{*}{$\Delta$} & \multirow{2}{*}{Model} 
        & \multicolumn{5}{c|}{NASDAQ100} 
        & \multicolumn{5}{c|}{S\&P500} & \multicolumn{5}{c}{NIFTY500} \\ \cline{3-17}
         &  
         & IRR & AIRR & SR & NDCG & ACC 
         & IRR & AIRR & SR & NDCG & ACC & IRR & AIRR & SR & NDCG & ACC \\ 
        \hline
        \hline

        \multirow{6}{*}{1} 
        & TRANSF 
        & 0.088 & 25.027 & 0.874 & 0.214 & 1.014  & 0.062 & 16.868 & 0.748 & 0.143 & 0.181 & 0.108 & 33.477 & 0.484 & 0.152 & 0.389 \\
        & STHGCN 
        & 0.054 & 14.654 & 0.761 & 0.209 & 0.861 & 0.054 & 14.841 & 0.667 & 0.139 & 0.194 & 0.055 & 14.982 & 0.620 & \textbf{0.210} & \textbf{0.820} \\
        & STHAN 
`       & 0.055 & 14.982 & 0.617 & 0.209 & 0.819 & 0.056 & 15.258 & 0.589 & 0.140 & 0.153 & 0.087 & 24.658 & -1.320 & 0.150 & 0.347 \\
        & GCNKG 
        & 0.090 & \textbf{27.302} & \textbf{0.880} & 0.213 & 1.056 & \textbf{0.100} & \textbf{27.840} & \textbf{0.920} & 0.142 & 0.292  & 0.128 & 38.140 & -0.621 & 0.148 & 0.278 \\
        & CI-STHPAN
        & 0.082 & 26.180 & 0.520 & 0.245 & 3.020 & 0.075 & 20.630 & 0.648 & 0.152 & 0.930 & 0.165 & 51.300 & 0.850 & 0.165 & 0.265 \\
        \rowcolor[HTML]{FFDAB9}
        & \textbf{TA-HKGE} 
        & \textbf{0.090} & 25.440 & 0.180 & \textbf{0.291} & \textbf{5.000} & 0.050 & 13.420 & 0.376 & \textbf{0.163} & \textbf{1.580} & \textbf{0.200} & \textbf{64.440} & \textbf{2.320} & 0.180 & 0.250 \\
        
        \hline
        \hline

        \multirow{5}{*}{5} & TRANSF
        & 0.403 & 22.524 & 1.787 & 0.214 & 1.125 & 0.247 & 13.263 & 1.307 & 0.139 & 0.042 & 0.457 & 25.848 & 1.360 & 0.147 & 0.250 \\
        & STHGCN
        & 0.455 & 25.744 & 1.926 & 0.215 & 1.306 & 0.323 & 17.640 & 1.542 & 0.142 & 0.306  & 0.498 & 28.956 & 1.149 & 0.149 & \textbf{0.472} \\
        & STHAN
        & 0.466 & 26.507 & \textbf{2.090} & 0.214 & 1.236 & \textbf{0.450} & \textbf{25.600} & \textbf{2.080} & 0.144 & 0.500 & 0.564 & 33.529 & -1.340 & 0.145 & 0.292 \\
        & GCNKG
        & 0.414 & 23.165 & 2.015 & 0.211 & 0.833  & 0.395 & 22.119 & 2.029 & 0.140 & 0.153 & 0.590 & 35.720 & 1.440 & 0.148 & 0.347 \\
        & CI-STHPAN
        & 0.580 & 34.200 & 1.420 & 0.255 & 3.420 & 0.382 & 21.200 & 1.620 & 0.155 & 0.820 & 0.690 & 42.380 & 1.880 & 0.150 & 0.358 \\ 
        \rowcolor[HTML]{FFDAB9}
        & \textbf{TA-HKGE}
        & \textbf{0.740} & \textbf{45.000} & 0.853 & \textbf{0.296} & \textbf{6.580} & 0.370 & 20.450 & 1.150 & \textbf{0.166} & \textbf{1.5}  & \textbf{0.800} & \textbf{49.420} & \textbf{2.290} & \textbf{0.153} & 0.370\\

        \hline
        \hline

        \multirow{5}{*}{20} & TRANSF
        & 1.641 & 22.782 & 3.780 & 0.210 & 1.264 & 1.029 & 13.769 & 2.892 & 0.139 & 0.153  & 2.152 & 30.789 & 3.369 & 0.147 & 0.264 \\
        & STHGCN
        & 1.702 & 23.719 & 4.370 & 0.207 & 1.069 & 1.685 & 23.617 & 3.964 & 0.143 & 0.653 & 2.633 & 38.794 & 4.092 & 0.150 & 0.920 \\
        & STHAN
        & 1.742 & 24.328 & 4.338 & 0.208 & 0.931 & \textbf{1.790} & \textbf{25.130} & 4.200 & 0.144 & 0.847 & 2.950 & 44.490 & 4.520 & 0.147 & 0.486\\
        & GCNKG
        & 1.504 & 21.345 & 3.848 & 0.211 & 1.417 & 1.172 & 16.001 & 3.196 & 0.138 & 0.181  & 2.632 & 38.978 & 4.109 & 0.147 & 0.569 \\
        & CI-STHPAN
        & 2.620 & 38.140 & 5.020 & 0.255 & 3.120 & 1.340 & 18.420 & 3.820 & 0.152 & 0.640 & 4.280 & 72.640 & 5.420 & 0.149 & 0.980 \\
        \rowcolor[HTML]{FFDAB9}
        & \textbf{TA-HKGE}
        & \textbf{3.510} & \textbf{54.460} & \textbf{6.067} & \textbf{0.289} & \textbf{4.870} & 1.500 & 20.630 & \textbf{4.400} & \textbf{0.166} & \textbf{1.080} & \textbf{6.650} & \textbf{125.060} & \textbf{6.585} & \textbf{0.150} & \textbf{1.420} \\

        \hline  
        \hline
    \end{tabular}
    \end{scriptsize}
\end{table}

\section{Additional Experiments and Discussion}
\label{appdx:sec:exps}

\begin{table}
  \caption{Best and worst case IRR across NASDAQ100, S\&P500, and NIFTY500 for different holding periods $\Delta$.}
  \label{ablation1}
  \begin{center}
    \begin{small}
      \begin{sc}
        \begin{tabular}{clcccc}
        \toprule
        \toprule
        $\Delta$ & Dataset & \multicolumn{2}{c}{Top 1 (IRR)} & \multicolumn{2}{c}{Top 5 (IRR)} \\
        \cmidrule(lr){3-4} \cmidrule(lr){5-6}
        ~ & ~ & Best & Worst & Best & Worst \\
        \midrule
        \midrule
        ~ & N100   & 4.06 & -3.65 & 3.67 & -3.26 \\
        1 & S\&P   & 2.98 & -2.79 & 2.65 & -2.41 \\
        ~ & NIFTY  & 5.12 & -3.41 & 3.77 & -3.43 \\
        \midrule
        \midrule
        ~ & N100 & 4.60 & -3.10 & 4.14 & -2.89 \\
        5 & S\&P   & 3.62 & -2.66 & 2.78 & -2.27 \\
        ~ & NIFTY  & 4.69 & -2.91 & 4.52 & -3.09 \\
        \midrule
        \midrule
        ~ & N100 & 4.26 & -3.22 & 4.80 & -2.21 \\
        20  & S\&P   & 4.68 & -1.58 & 4.33 & -1.12 \\
        ~  & NIFTY  & 6.13 & -2.56 & 5.61 & -2.32 \\
        \bottomrule
        \bottomrule
      \end{tabular}
      \end{sc}
    \end{small}
  \end{center}
  \vskip -0.1in
\end{table}

\paragraph{What is the best and worst case scenario in terms of return on investment?}

In practice, it is difficult to execute transactions optimally on a given day. To account for this uncertainty,we evaluated the best- and worst-case scenarios in terms of the investment return ratio (IRR). The best case scenario is when the stock is bought at its lowest price on day $T$ and sold at its highest price on day $T + \Delta$, whereas the worst case scenario is when the stock is bought at its highest price on day $T$ and sold at its lowest price on day $T + \Delta$. Table \ref{ablation1} shows the best and worst case IRR obtained across NASDAQ100, S\&P500, and NIFTY500 using our model. The variance in the best and worst IRR decreases as the holding period increases. In addition, the results highlight the importance of accurately predicting closing prices is important task for real-world applications of this work.

\begin{table*}
  \caption{Standard deviation of metrics across all 24 phases using TA-HKGE.}
  \label{table:standard_deviation}
  \begin{center}
    \begin{small}
      \begin{sc}
        \begin{tabular}{l l llll|llll }
        \hline
        \hline
        \multirow{2}{*}{$\Delta$} & \multirow{2}{*}{Dataset} & \multicolumn{4}{c|}{Top 1} &\multicolumn{4}{c}{Top 5} \\ \cline{3-10}
         &  & NDCG & ACC & IRR & SR & NDCG & ACC & IRR & SR \\ 
        \hline
        \hline
        \multirow{3}{*}{1}  
        & \cellcolor{White} N100   & \cellcolor{White} 0.037 & \cellcolor{White} 3.24 & \cellcolor{White} 0.27 & \cellcolor{White} 1.62 & \cellcolor{White} 0.032 & \cellcolor{White} 3.22 & \cellcolor{White} 0.19 & \cellcolor{White} 0.35 \\ 
        & \cellcolor{Grey1} S\&P  & \cellcolor{Grey1} 0.029 & \cellcolor{Grey1} 2.58 & \cellcolor{Grey1} 0.23 & \cellcolor{Grey1} 1.76 & \cellcolor{Grey1} 0.023 & \cellcolor{Grey1} 1.6 & \cellcolor{Grey1} 0.15 & \cellcolor{Grey1} 0.41 \\ 
        & \cellcolor{Grey2} NIFTY  & \cellcolor{Grey2} 0.008 & \cellcolor{Grey2} 0.49 & \cellcolor{Grey2} 0.30 & \cellcolor{Grey2} 2.86 & \cellcolor{Grey2} 0.006  & \cellcolor{Grey2} 0.84 & \cellcolor{Grey2} 0.17  & \cellcolor{Grey2} 0.41 \\ 
        \hline
        \hline
        
        \multirow{3}{*}{5}  
        & \cellcolor{White} N100   & \cellcolor{White} 0.05 & \cellcolor{White} 6.40 & \cellcolor{White} 1.42 & \cellcolor{White} 4.23 & \cellcolor{White} 0.032 & \cellcolor{White} 3.94 & \cellcolor{White} 0.84 & \cellcolor{White} 0.69 \\ 
        & \cellcolor{Grey1} S\&P  & \cellcolor{Grey1} 0.022 & \cellcolor{Grey1} 2.39 & \cellcolor{Grey1} 1.11 & \cellcolor{Grey1} 3.37 & \cellcolor{Grey1} 0.018 & \cellcolor{Grey1} 1.42 & \cellcolor{Grey1} 0.85 & \cellcolor{Grey1} 1.07 \\ 
        & \cellcolor{Grey2} NIFTY  & \cellcolor{Grey2} 0.023 & \cellcolor{Grey2} 1.79 & \cellcolor{Grey2} 1.36 & \cellcolor{Grey2} 3.42 & \cellcolor{Grey2} 0.02 & \cellcolor{Grey2} 2.24 & \cellcolor{Grey2} 0.96 & \cellcolor{Grey2} 0.86 \\ \hline
        \hline

        \multirow{3}{*}{20} 
        & \cellcolor{White} N100   & \cellcolor{White} 0.05 & \cellcolor{White} 4.66 & \cellcolor{White} 5.63 & \cellcolor{White} 8.26 & \cellcolor{White} 0.047 & \cellcolor{White} 5.39 & \cellcolor{White} 4.25 & \cellcolor{White} 3.08 \\ 
        & \cellcolor{Grey1} S\&P  & \cellcolor{Grey1} 0.035 & \cellcolor{Grey1} 3.01 & \cellcolor{Grey1} 3.42 & \cellcolor{Grey1} 13.89 & \cellcolor{Grey1} 0.032 & \cellcolor{Grey1} 3.27 & \cellcolor{Grey1} 2.25 & \cellcolor{Grey1} 1.94 \\ 
        & \cellcolor{Grey2} NIFTY  & \cellcolor{Grey2} 0.019 & \cellcolor{Grey2} 1.41 & \cellcolor{Grey2} 6.34 & \cellcolor{Grey2} 11.56 & \cellcolor{Grey2} 0.05 & \cellcolor{Grey2} 2.16 & \cellcolor{Grey2} 3.98 & \cellcolor{Grey2} 2.53 \\ \hline
        \hline
    \end{tabular}
      \end{sc}
    \end{small}
  \end{center}
  \vskip -0.1in
\end{table*}

\noindent
\textbf{What is the standard deviation of metrics across the 24 phases?}
Market conditions vary significantly across time, with different phases exhibiting bullish or bearish behavior. To assess the robustness of the proposed method under such variability, we analyze the standard deviations of the evaluation metrics across all 24 market phases. Table \ref{table:standard_deviation} shows the standard deviation of the metrics across all 24 phases obtained using our model. The standard deviation is comparatively lower for larger datasets, such as the S\&P500 and NIFTY500. Furthermore, for all datasets and holding periods, selecting a larger set of stocks (Top 5) results in lower variability than selecting a single stock (Top 1). This suggests that portfolio diversification, together with the proposed temporal knowledge graph model, contributes to a more consistent performance across different market conditions.

\noindent
\textbf{Performance on the Full Dataset} 
In Section~\ref{section:Experiemental_Setup}, we have provided the results applying our method and baselines models on subsets of the whole dataset obtained as a sliding window of 24 "phases" and averaging out the results over all the phases. To study the effect of learning over a long time horizon, we provide the results of running the algorithms by dividing the entire dataset into training and test sets in Tables~\ref{nasdaq_fulldataset}, \ref{sp_fulldataset}, and \ref{nifty_fulldataset}.

\noindent
\textbf{Effect of Model Components on Performance}
To analyze the contribution of individual components in TA-HKGE. We empirically compared multiple variants of our model, TA-HKGE. For the first variant, WOTPP, we removed the embeddings obtained from temporal process modeling and only considered sequential and relational embeddings. For the second variant, LSTM*, we replace the transformer-based sequential embedding in WOTPP with an LSTM. Next, we consider WOSEQ, in which the sequential embedding update for the stock assets as input for the relational embedding layer is removed, along with the embeddings obtained from the temporal point process model. Finally, the temporal point process embeddings and the heterogeneous transformation of entity features are removed to obtain WOHK. Table ~\ref{change} shows that each component of our architecture plays an important role in improving the performance of the algorithm.

{
\setlength{\tabcolsep}{5.2pt} 
\begin{table*}[]
    \centering
    \caption{Comparison of TA-HKGE with baseline methods on the NASDAQ100 dataset under full-dataset training. Results across multiple holding periods are reported, with the best values highlighted in \textbf{bold}. TA-HKGE outperforms baseline models on the majority of evaluation metrics,.}
    \label{nasdaq_fulldataset}
    \begin{scriptsize}
    \begin{tabular}{l p{1.3cm} ccccc | ccccc}
        \hline
        \hline
        \multirow{2}{*}{$\Delta$} & \multirow{2}{*}{Model} 
        & \multicolumn{5}{c|}{Top 1} 
        & \multicolumn{5}{c}{Top 5} \\ \cline{3-12}
         &  
         & IRR & AIRR & SR & NDCG & ACC 
         & IRR & AIRR & SR & NDCG & ACC \\ 
        \hline
        \hline
        \multirow{5}{*}{1} 
        & TRANSF 
        & -30 & -58.5 & -1.322 & 0.2646 & 1.81 & -9.0 & -26.6 & -0.2002 & 0.3771 & 10.69 \\
        & STHAN 
        & -11.8 & -26.9 & -0.5448 & 0.2664 & 3.24 & -14.2 & -30.2 & -0.2883 & 0.3673 & 10.97 \\
        & GCNKG 
        & -12.5 & -28.3 & -0.4952 & 0.2773 & 5.05 & \textbf{-0.16} & -30.2 & \textbf{-0.2835} & 0.3799 & \textbf{11.19} \\
        & CI-STHPAN
        & -8.90 & -19.40 & -0.3200 & 0.2801 & 5.82 & -4.60 & -25.10 & -0.2550 & 0.3750 & 10.05 \\
        \rowcolor[HTML]{FFDAB9}
        &\textbf{TA-HKGE} 
        & \textbf{-4.80} & \textbf{-11.6} & \textbf{-0.1845} & \textbf{0.283} & \textbf{6.49} & -8.81 & \textbf{-20.6} & -0.226 & \textbf{0.3712} & 9.02 \\

        \hline
        \hline

        \multirow{5}{*}{5} & TRANSF 
        & -0.63 & -79.66 & -1.206 & 0.2601 & 2.53 & -0.61 & -78.6 & -0.4334 & 0.3677 & 7.87 \\
        & STHAN 
        &  -0.82 & -87.57 & -1.4353 & 0.2865 & 5.77 & -0.98 & -90.63 & -0.5805 & 0.3843 & \textbf{11.26} \\
        & GCNKG 
        & -1.4 & -97.11 & -2.436 & 0.3072 & \textbf{6.13} & -0.65 & -69.09 & -0.4497 & \textbf{0.3935} & 12.7 \\
        & CI-STHPAN
        & 10.20 & -30.50 & -0.9000 & 0.2980 & 4.80 & 2.30 & -20.00 & -0.2200 & 0.3600 & 8.60 \\
        \rowcolor[HTML]{FFDAB9}
        & \textbf{TA-HKGE} 
        &  \textbf{23.14} & \textbf{11.9} & \textbf{0.6226} & \textbf{0.2899} & 3.61 & \textbf{6.24} & \textbf{3.2} & \textbf{0.0075} & 0.3227 & 4.98 \\
        
        \hline
        \hline

        \multirow{5}{*}{20} & TRANSF 
        & -2.84 & -99.93 & -2.4777 & 0.2522 & 2.89 & -2.45 & -99.9 & -0.868 & 0.3834 & \textbf{14.01} \\
        & STHAN 
        & -3.09 & -99.96 & -2.823 & 0.2422 & 1.8 & -3.43 & -99.9 & -1.112 & 0.3782 & 12.56 \\
        & GCNKG 
        & -1.90 & -90.23 & \textbf{-0.9613} & 0.233 & 2.52 & -3.38 & -99.9 & -1.150 & 0.3745 & 12.49 \\
        & CI-STHPAN
        & -1.885 & -55.40 & -1.6000 & 0.2550 & 4.50 & -2.50 & -55.60 & -0.9500 & 0.3800 & 11.20 \\
        \rowcolor[HTML]{FFDAB9}
        &\textbf{TA-HKGE} & 
        \textbf{-1.87} & \textbf{-21.6} & \textbf{-2.5309} & \textbf{0.2796} & \textbf{6.85} & \textbf{-1.83} & \textbf{-21.0} & \textbf{-0.7427} & \textbf{0.3866} & 10.1 \\

        \hline  
        \hline
    \end{tabular}
    \end{scriptsize}
\end{table*}

\begin{table*}[]
    \centering
    \caption{Comparison of TA-HKGE with baseline methods on the S\&P500 dataset under full-dataset training. Results across multiple holding periods are reported, with the best values highlighted in \textbf{bold}. TA-HKGE outperforms baseline models on the majority of evaluation metrics,}
    \label{sp_fulldataset}
    \begin{scriptsize}
    \begin{tabular}{l p{1.3cm} ccccc | ccccc}
        \hline
        \hline
        \multirow{2}{*}{$\Delta$} & \multirow{2}{*}{Model} 
        & \multicolumn{5}{c|}{Top 1} 
        & \multicolumn{5}{c}{Top 5} \\ \cline{3-12}
         &  
         & IRR & AIRR & SR & NDCG & ACC 
         & IRR & AIRR & SR & NDCG & ACC \\ 
        \hline
        \hline
        \multirow{6}{*}{1} 

        & TRANSF
        & \textbf{-0.15} & -31.5 & -0.6494 & \textbf{0.1599} & 0.72 & -0.14 & -24.79 & -0.301 & 0.233 & \textbf{2.53} \\
        & STHAN
        & -0.24 & -45.42 & -1.4418 & 0.1558 & 0.36 & -0.05 & -12.17 & -0.2117 & 0.2186 & 1.73 \\
        & GCNKG
        & -0.42 & -65.2 & -2.0955 & 0.1586 & \textbf{1.08} & -0.09 & -13.95 & -0.2976 & 0.225 & 2.02 \\
        & CI-STHPAN
        & -0.425 & -30.00 & -1.1000 & 0.1500 & 0.600 & 3.50 & 5.80 & -0.1200 & 0.2170 & 1.20 \\
        \rowcolor[HTML]{FFDAB9}
        & \textbf{TA-HKGE}
        & -0.43 & \textbf{-1.08} & \textbf{-0.0618} & 0.1429 & 0 & \textbf{7.20} & \textbf{20.2} & \textbf{0.0443} & \textbf{0.2087} & 0.72 \\

        \hline
        \hline

        \multirow{5}{*}{5} 
        & TRANSF
        & -0.41 & -64.49 & -1.3532 & 0.1501 & 0 & -0.49 & -57.03 & -0.5134 & 0.2196 & 1.52 \\
        & STHAN
        & -0.61 & -78.6 & -1.4651 & 0.1518 & 0 & -0.04 & -9.71 & -0.0767 & \textbf{0.2234} & \textbf{2.45} \\
        & GCNKG
        & -0.51 & -72.43 & -1.356 & 0.141 & 0 & -0.32 & -41.61 & -0.3208 & 0.2159 & 1.66 \\
        & CI-STHPAN
        & 3.20 & -34.10 & -0.5400 & 0.1470 & 0 & 12.40 & 23.80 & 0.0100 & 0.2120 & 1.30 \\
        \rowcolor[HTML]{FFDAB9}
        & \textbf{TA-HKGE}
        & \textbf{7.51} & \textbf{3.9} & \textbf{0.2717} & \textbf{0.1529} & \textbf{0} & \textbf{34.2} & \textbf{89.3} & \textbf{0.3642} & 0.2087 & 1.01 \\

        \hline
        \hline

        \multirow{5}{*}{20}
        & TRANSF
        & -1.10 & -94.15 & -1.8781 & 0.1457 & 0 & -1.34 & -95.41 & -0.6883 & 0.2133 & 1.37 \\
        & STHAN
        & -1.5 & -98.05 & -2.666 & 0.147 & 0.36 & -0.51 & -65.65 & -0.2844 & 0.2095 & \textbf{1.81} \\
        & GCNKG
        & -0.09 & -20.3 & -1.1363 & 0.1553 & 1.44 & \textbf{-0.24} & -45.42 & \textbf{-0.1537} & \textbf{0.2119} & 2.09 \\
        & CI-STHPAN
        & 0.210 & -12.80 & -1.0800 & 0.1560 & 0.720 & -0.980 & -12.60 & -0.7200 & 0.2060 & 1.05 \\
        \rowcolor[HTML]{FFDAB9}
        & \textbf{TA-HKGE}
        & \textbf{0.496} & \textbf{-6.1} & \textbf{-1.0253} & \textbf{0.1572} & \textbf{0} & -1.71 & \textbf{19.1} & -1.2973 & 0.2006 & 0 \\

        \hline
        \hline
        
    \end{tabular}
    \end{scriptsize}
\end{table*}

\begin{table*}[]
    \centering
    \caption{Comparison of TA-HKGE with baseline methods on the NIFTY500 dataset under full-dataset training. Results across multiple holding periods are reported, with the best values highlighted in \textbf{bold}. TA-HKGE outperforms baseline models on the majority of evaluation metrics,}
    \label{nifty_fulldataset}
    \begin{scriptsize}
    \begin{tabular}{l p{1.3cm} ccccc | ccccc}
        \hline
        \hline
        \multirow{2}{*}{$\Delta$} & \multirow{2}{*}{Model} 
        & \multicolumn{5}{c|}{Top 1} 
        & \multicolumn{5}{c}{Top 5} \\ \cline{3-12}
         &  
         & IRR & AIRR & SR & NDCG & ACC 
         & IRR & AIRR & SR & NDCG & ACC \\ 
        \hline
        \hline
        \multirow{6}{*}{1} 

        & TRANSF
        & -0.18 & -36.49 & -1.3665 & 0.1544 & 0 & -0.02 & -4.86 & -1.3643 & 0.2292 & 1.52 \\
        & STHAN
        & -0.01 & -34.8 & -0.3998 & 0.1621 & 0 & 0.16 & 48.92 & -1.0408 & \textbf{0.2345} & \textbf{2.45} \\
        & GCNKG
        & 0.01 & 2.60 & -0.1551 & 0.1511 & 0 & 0.06 & -1.61 & -1.515 & 0.2237 & 1.29 \\
        & CI-STHPAN
        & 3.40 & 11.20 & 0.1200 & 0.1515 & 0 & 2.10 & 6.40 & -1.4700 & 0.2237 & 1.26 \\
        \rowcolor[HTML]{FFDAB9}
        & \textbf{TA-HKGE}
        & \textbf{7.0} & \textbf{19.5} & \textbf{0.3843} & \textbf{0.1518} & \textbf{0} & \textbf{4.89} & \textbf{13.5} & \textbf{-1.4227} & 0.2237 & 1.23 \\

        \hline
        \hline

        \multirow{5}{*}{5} 
        & TRANSF
        & -0.62 & -79.14 & -1.7195 & 0.1493 & 0 & -0.29 & -72.21 & -0.9045 & \textbf{0.2232} & 1.52 \\
        & STHAN
        & -0.91 & -91.42 & -2.761 & 0.1527 & 0 & \textbf{0.06} & 17.12 & \textbf{-0.5584} & 0.2239 & \textbf{2.23} \\
        & GCNKG
        & \textbf{0.04} & \textbf{-11.04} & \textbf{0.0318} & 0.1418 & 0 & -0.13 & -39.85 & -0.7612 & 0.2127 & 0.86 \\
        & CI-STHPAN
        & -13.50 & -28.40 & -0.6890 & 0.1465 & 0 & -10.15 & -39.82 & -0.9750 & 0.2170 & 0.54 \\
        \rowcolor[HTML]{FFDAB9}
        & \textbf{TA-HKGE}
        & -27.01 & -45.7 & -1.4099 & \textbf{0.1513} & \textbf{0} & -20.17 & \textbf{-39.8} & -1.1897 & 0.2210 & 0.21 \\

        \hline
        \hline

        \multirow{5}{*}{20}
        & TRANSF
        & -2.78 & -99.9 & -2.9371 & 0.1271 & 0 & -1.10 & -94.14 & -1.2607 & 0.1981 & \textbf{0.87} \\
        & STHAN
        & -3.66 & -99.9 & -5.645 & 0.1439 & 0 & -0.86 & -90.23 & -0.729 & 0.2126 & 0.79 \\
        & GCNKG
        & -0.55 & -74.92 & -1.124 & 0.1298 & 0 & -1.05 & -93.66 & -1.131 & 0.1984 & 0.43 \\
        & CI-STHPAN
        & -0.525 & -66.85 & -0.9870 & 0.1406 & 0 & -0.91 & -82.63 & -1.0220 & 0.2100 & 0.25 \\
        \rowcolor[HTML]{FFDAB9}
        & \textbf{TA-HKGE}
        & \textbf{-0.5} & \textbf{-58.8} & \textbf{-0.8499} & \textbf{0.1514} & \textbf{0} & \textbf{-0.77} & \textbf{-71.6} & \textbf{-0.914} & \textbf{0.2216} & 0.07 \\

        \hline
        \hline
        
    \end{tabular}
    \end{scriptsize}
\end{table*}
}

\noindent
\textbf{Effect of Increasing Training, Validation, and Testing Time in each phase on Model Performance} 
As the stock price drifts in a direction over a period of time, it can negatively affect model training. Especially when the training and testing periods are sufficiently long, the stock price and the prevailing socio-political and economic conditions might have changed from the training to the test period. To measure the extent of this effect, two more training setups are considered, the first of which (750TR) has a training period of 750 trading days (3 years), a 75-day validation period, and a 200-day testing period. In the next setup (1000TR), the training period is 1000 trading days (4 years), 100-day validation, and 300-day testing period in each of the 21 phases. As shown in Table \ref{days_increase} the performance is affected when the training period is increased. However, the results obtained from the model still beat risk-free returns as indicated by the Sharpe ratio.

\subsection{Hyperparameters}
\label{appdx:sec:parameters}
The important hyperparameters used for the algorithm are listed in Table~\ref{hyperparameters}. 
The weight coefficients $\alpha_i$ for each loss component in the objective were set to $1$.
We performed experiments by varying the size of the Relational Embedding in $\{ 16,32,64,128 \}$ and chose the best-performing dimension. We set the size of the sequential embeddings to 20 and that of the relation embeddings to 16 for NIFTY and 128 for NASDAQ and S\&P. For ablation studies, we used embedding sizes of 20 and 16, respectively, to evaluate their effect on model performance.

\begin{table}
  \caption{Hyperparameters}
  \label{hyperparameters}
  \begin{center}
    \begin{small}
      \begin{sc}
        \begin{tabular}{ll}
        \toprule
        \toprule
        Parameter & Value \\ 
        \midrule
        \midrule
        Learning Rate & 0.00001 \\ 
        Number of epochs & 10 \\ 
        Temporal Point Process embeddings size & 128 \\ 
        Learning rate for HPGE model & 0.0001 \\  
        \bottomrule
        \bottomrule
        \end{tabular}
      \end{sc}
    \end{small}
  \end{center}
  \vskip -0.1in
\end{table}

\begin{table*}[t]
\centering
\caption{Ablation analysis: comparison of TA-HKGE without various components of the architecture as described in Section~\ref{sec:ablations}.}
\label{change}
\begin{scriptsize}
\begin{tabular}{l l l lllll|lllll}
\hline
\hline
    
    \multirow{2}{*}{$\Delta$} & \multirow{2}{*}{Model} & \multirow{2}{*}{Dataset} & \multicolumn{5}{c|}{Top 1} &\multicolumn{5}{c}{Top 5} \\ \cline{4-13}
    &  &  & IRR & AIRR & SR & NDCG & ACC & IRR & AIRR & SR & NDCG & ACC \\ \hline
    \hline
     
    \multirow{12}{*}{1} & \multirow{3}{*}{WOSEQ} & \cellcolor{White} N100 & \cellcolor{White} {0.13} & \cellcolor{White} {38.73} & \cellcolor{White} {0.60} & \cellcolor{White} {0.27} & \cellcolor{White} {3.87} & \cellcolor{White} {0.08} & \cellcolor{White} {22.32} & \cellcolor{White} {0.04} & \cellcolor{White} {0.37} & \cellcolor{White} {10.82} \\ \cline{3-13}
     & ~ & \cellcolor{Grey1} S\&P & \cellcolor{Grey1} 0.103 & \cellcolor{Grey1} 19.61 & \cellcolor{Grey1} 0.77 & \cellcolor{Grey1} {0.15} & \cellcolor{Grey1} {0.49} & \cellcolor{Grey1} {0.08} & \cellcolor{Grey1} {22.32} & \cellcolor{Grey1} {0.05} & \cellcolor{Grey1} {0.221} & \cellcolor{Grey1} {1.95} \\ \cline{3-13}
     & ~ &  \cellcolor{Grey2} NIFTY & \cellcolor{Grey2} 0.15 & \cellcolor{Grey2} 45.89 & \cellcolor{Grey2} 0.85 & \cellcolor{Grey2} 0.154 & \cellcolor{Grey2} 0.29 & \cellcolor{Grey2} {0.14} & \cellcolor{Grey2} {42.26} & \cellcolor{Grey2} 0.244 & \cellcolor{Grey2} 0.225 & \cellcolor{Grey2} 1.67 \\ \cline{2-13}

    &  \multirow{3}{*}{WOHK} & \cellcolor{White} N100 & \cellcolor{White} {0.13} & \cellcolor{White} {38.73} & \cellcolor{White} {0.61}& \cellcolor{White} {0.29} & \cellcolor{White} {5.24} & \cellcolor{White} {0.13} & \cellcolor{White} {38.73} & \cellcolor{White} {0.14} & \cellcolor{White} {0.40} & \cellcolor{White} {14.4}  \\ \cline{3-13}
     & ~ & \cellcolor{Grey1} S\&P & \cellcolor{Grey1} {0.59} & \cellcolor{Grey1} {340.36} & \cellcolor{Grey1} 0.23 & \cellcolor{Grey1} {0.17} & \cellcolor{Grey1} {1.37} & \cellcolor{Grey1} {0.10} & \cellcolor{Grey1} {28.64} & \cellcolor{Grey1} {0.078} & \cellcolor{Grey1} {0.24} & \cellcolor{Grey1} {3.44} \\ \cline{3-13}
     & ~ & \cellcolor{Grey2} NIFTY & \cellcolor{Grey2} {0.06} & \cellcolor{Grey2} {16.31} & \cellcolor{Grey2} 0.425 & \cellcolor{Grey2} {0.157} & \cellcolor{Grey2} {0.33} & \cellcolor{Grey2} {0.13} & \cellcolor{Grey2} {38.73} & \cellcolor{Grey2} {0.22} & \cellcolor{Grey2} 0.22 & \cellcolor{Grey2} {2.08} \\ \cline{2-13}

    & \multirow{3}{*}{LSTM*} & \cellcolor{White} N100 & \cellcolor{White} {0.16} & \cellcolor{White} {49.61} & \cellcolor{White} {0.77} & \cellcolor{White} {0.28}& \cellcolor{White} {5.12} & \cellcolor{White} {0.13} & \cellcolor{White} {38.73} & \cellcolor{White} {0.14} & \cellcolor{White} {0.39} & \cellcolor{White} {12.54} \\ \cline{3-13}
     & ~ & \cellcolor{Grey1} S\&P & \cellcolor{Grey1} 0.016 & \cellcolor{Grey1} 4.11 & \cellcolor{Grey1} {0.279} & \cellcolor{Grey1} {0.16} & \cellcolor{Grey1} 0.62 &  \cellcolor{Grey1} 0.06 & \cellcolor{Grey1} {16.31} & \cellcolor{Grey1} 0.031 & \cellcolor{Grey1} {0.23}& \cellcolor{Grey1} {2.57} \\ \cline{3-13}
     & ~ & \cellcolor{Grey2} NIFTY & \cellcolor{Grey2} {0.17} & \cellcolor{Grey2} {53.42} & \cellcolor{Grey2} {1.00} & \cellcolor{Grey2} 0.15 & \cellcolor{Grey2} {0.58} & \cellcolor{Grey2} {0.12} & \cellcolor{Grey2} {35.28} & \cellcolor{Grey2} 0.17 & \cellcolor{Grey2} 0.22 & \cellcolor{Grey2} 1.76 \\ \cline{2-13}
         
    & \multirow{3}{*}{ WOTPP } & \cellcolor{White} N100 & \cellcolor{White} {0.09} & \cellcolor{White} 25.44 &  \cellcolor{White} 0.48 & \cellcolor{White} {0.27} & \cellcolor{White} {3.58} & \cellcolor{White} {0.11} & \cellcolor{White} {31.92} & \cellcolor{White} {0.11} & \cellcolor{White} {0.37} & \cellcolor{White} {10.77} \\ \cline{3-13}
     & ~ & \cellcolor{Grey1} S\&P & \cellcolor{Grey1} {0.11} & \cellcolor{Grey1} {31.92} & \cellcolor{Grey1} 0.90  & \cellcolor{Grey1} {0.16} & \cellcolor{Grey1} {0.79} & \cellcolor{Grey1} 0.09 & \cellcolor{Grey1} {25.44} & \cellcolor{Grey1} 0.99 & \cellcolor{Grey1} {0.22} & \cellcolor{Grey1} {2.23} \\ \cline{3-13}
     & ~ &  \cellcolor{Grey2} NIFTY & \cellcolor{Grey2} {0.18} & \cellcolor{Grey2} {57.33} & \cellcolor{Grey2} {1.14} & \cellcolor{Grey2} {0.156}& \cellcolor{Grey2} {0.29}& \cellcolor{Grey2} {0.13}& \cellcolor{Grey2} {38.73} & \cellcolor{Grey2} {0.19} & \cellcolor{Grey2} {0.23} &  \cellcolor{Grey2}{1.65} \\

     \hline
     \hline
     
    \multirow{12}{*}{5} & \multirow{3}{*}{WOSEQ} & \cellcolor{White} N100 & \cellcolor{White} {0.55} & \cellcolor{White} {31.84} & \cellcolor{White} {1.38} & \cellcolor{White} {0.27} & \cellcolor{White} {3.87} & \cellcolor{White} {0.58} & \cellcolor{White} {33.84} & \cellcolor{White} {0.54} & \cellcolor{White} {0.38} & \cellcolor{White} {13.39} \\ \cline{3-13}
     & ~ & \cellcolor{Grey1} S\&P & \cellcolor{Grey1} 0.06 & \cellcolor{Grey1} 3.06 & \cellcolor{Grey1} 0.56 & \cellcolor{Grey1} {0.16} & \cellcolor{Grey1} {1.20} & \cellcolor{Grey1} {0.236} & \cellcolor{Grey1} {12.61} & \cellcolor{Grey1} {0.34} & \cellcolor{Grey1} {0.22} & \cellcolor{Grey1} {2.36} \\ \cline{3-13}
     & ~ &  \cellcolor{Grey2} NIFTY & \cellcolor{Grey2} 0.55 & \cellcolor{Grey2} 31.84 & \cellcolor{Grey2} 0.86 & \cellcolor{Grey2} {0.151} & \cellcolor{Grey2} {0.24} & \cellcolor{Grey2} {0.39}& \cellcolor{Grey2} {21.67} & \cellcolor{Grey2} {0.53} & \cellcolor{Grey2} {0.227} & \cellcolor{Grey2} {1.56} \\ \cline{2-13}
      
    & \multirow{3}{*}{WOHK} & \cellcolor{White} N100 & \cellcolor{White} {0.50} & \cellcolor{White} 28.57 & \cellcolor{White} {0.23} & \cellcolor{White} 0.31 & \cellcolor{White} 6.02 & \cellcolor{White} 0.68 & \cellcolor{White} 40.71 & \cellcolor{White} {0.57} & \cellcolor{White} 0.41 & \cellcolor{White} 15.8 \\ \cline{3-13}
     & ~ & \cellcolor{Grey1} S\&P & \cellcolor{Grey1} 0.624 & \cellcolor{Grey1} 36.82 & \cellcolor{Grey1} 1.90 & \cellcolor{Grey1} 0.17 & \cellcolor{Grey1} 1.66 & \cellcolor{Grey1} 0.417 & \cellcolor{Grey1} 23.33 & \cellcolor{Grey1} 0.49 & \cellcolor{Grey1} 0.24 & \cellcolor{Grey1} 4.14 \\ \cline{3-13}
     & ~ & \cellcolor{Grey2} NIFTY & \cellcolor{Grey2} {0.745} & \cellcolor{Grey2} {45.36} & \cellcolor{Grey2} {1.86} & \cellcolor{Grey2} 0.16 & \cellcolor{Grey2} 0.79 & \cellcolor{Grey2} 0.63 & \cellcolor{Grey2} 37.23 & \cellcolor{Grey2} 0.646 & \cellcolor{Grey2} 0.23 & \cellcolor{Grey2} 2.65 \\ \cline{2-13}

    & \multirow{3}{*}{LSTM*} & \cellcolor{White} N100 & \cellcolor{White} {0.68} & \cellcolor{White} {40.71} & \cellcolor{White} 1.51 & \cellcolor{White} {0.28} & \cellcolor{White} 4.83 & \cellcolor{White} {0.5} & \cellcolor{White} {28.57} & \cellcolor{White} {0.49} & \cellcolor{White} {0.38} & \cellcolor{White} {12.6} \\ \cline{3-13}
     & ~ & \cellcolor{Grey1} S\&P & \cellcolor{Grey1} {0.68} & \cellcolor{Grey1} {40.71} & \cellcolor{Grey1} {1.64} & \cellcolor{Grey1} 0.17 & \cellcolor{Grey1} 2.08 & \cellcolor{Grey1} {0.36} & \cellcolor{Grey1} {19.85} & \cellcolor{Grey1} {0.36} & \cellcolor{Grey1} 0.24 & \cellcolor{Grey1} 3.74 \\ \cline{3-13}
     & ~ &  \cellcolor{Grey2} NIFTY & \cellcolor{Grey2} 0.72 & \cellcolor{Grey2} 43.55 & \cellcolor{Grey2} 1.45 & \cellcolor{Grey2} {0.15} & \cellcolor{Grey2} 0.99& \cellcolor{Grey2} 0.72& \cellcolor{Grey2} 43.55 & \cellcolor{Grey2} 0.76 & \cellcolor{Grey2} {0.22} & \cellcolor{Grey2} 2.10 \\ \cline{2-13}
         
    & \multirow{3}{*}{WOTPP} & \cellcolor{White} N100 & \cellcolor{White} 0.71 & \cellcolor{White} 42.84 & \cellcolor{White} 2.18 & \cellcolor{White} 0.28 & \cellcolor{White} {4.24} & \cellcolor{White} 0.6 & \cellcolor{White} 35.18 & \cellcolor{White} 0.586 & \cellcolor{White} 0.38 & \cellcolor{White} 12.7 \\ \cline{3-13}
     & ~ & \cellcolor{Grey1} S\&P & \cellcolor{Grey1} 0.77 & \cellcolor{Grey1} 47.19 & \cellcolor{Grey1} 3.07 & \cellcolor{Grey1} {0.16} & \cellcolor{Grey1} {0.95} & \cellcolor{Grey1} 0.55 & \cellcolor{Grey1} 31.84 & \cellcolor{Grey1} 0.68 & \cellcolor{Grey1} {0.22} & \cellcolor{Grey1} {2.51} \\ \cline{3-13}
     & ~ &  \cellcolor{Grey2} NIFTY & \cellcolor{Grey2} 0.98 & \cellcolor{Grey2} 63.47 & \cellcolor{Grey2} 2.32 & \cellcolor{Grey2} 0.15& \cellcolor{Grey2} {0.54}& \cellcolor{Grey2} 0.62 & \cellcolor{Grey2} 36.54 & \cellcolor{Grey2} 0.56 & \cellcolor{Grey2} 0.22 & \cellcolor{Grey2}1.61 \\

   \hline
   \hline
   
    \multirow{12}{*}{20} & \multirow{3}{*}{WOSEQ} & \cellcolor{White} N100 & \cellcolor{White} 3.35 & \cellcolor{White} 51.46 & \cellcolor{White} 5.22 & \cellcolor{White} 0.29 & \cellcolor{White} 7.83 & \cellcolor{White} 1.92 & \cellcolor{White} 27.07 & \cellcolor{White} 0.86 & \cellcolor{White} 0.39 & \cellcolor{White} 13.36 \\ \cline{3-13}
     & ~ & \cellcolor{Grey1} S\&P & \cellcolor{Grey1} 1.68 & \cellcolor{Grey1} 23.35 & \cellcolor{Grey1} 0.45 & \cellcolor{Grey1} 0.177 & \cellcolor{Grey1} 2.37 & \cellcolor{Grey1} 1.36 & \cellcolor{Grey1} 18.55 & \cellcolor{Grey1} 0.89 & \cellcolor{Grey1} 0.24 & \cellcolor{Grey1} 4.10  \\ \cline{3-13}
     & ~ & \cellcolor{Grey2} NIFTY & \cellcolor{Grey2} 3.53 & \cellcolor{Grey2} 54.82 & \cellcolor{Grey2} 2.57 & \cellcolor{Grey2} 0.153 & \cellcolor{Grey2} 0.66 & \cellcolor{Grey2} 2.48 & \cellcolor{Grey2} 36.16 & \cellcolor{Grey2} 1.37 & \cellcolor{Grey2} 0.229 & \cellcolor{Grey2} 2.07 \\ \cline{2-13}
      
    & \multirow{3}{*}{WOHK} & \cellcolor{White} N100 & \cellcolor{White} {0.84} & \cellcolor{White} {36.16} & \cellcolor{White} 0.617 & \cellcolor{White} {0.28} & \cellcolor{White} {3.33} & \cellcolor{White} {1.43} & \cellcolor{White} {19.59}& \cellcolor{White} {0.616} & \cellcolor{White} 0.39 & \cellcolor{White} 13.98  \\ \cline{3-13}
     & ~ & \cellcolor{Grey1} S\&P & \cellcolor{Grey1} 0.14 & \cellcolor{Grey1} 1.77 & \cellcolor{Grey1} -3.11 & \cellcolor{Grey1} 0.179 & \cellcolor{Grey1} 1.87 & \cellcolor{Grey1} 0.92 & \cellcolor{Grey1} 12.23 & \cellcolor{Grey1} 0.16 & \cellcolor{Grey1} 0.24 & \cellcolor{Grey1} 4.77 \\ \cline{3-13}
     & ~ &  \cellcolor{Grey2} NIFTY & \cellcolor{Grey2} {2.48} & \cellcolor{Grey2} {36.16} & \cellcolor{Grey2} {2.96} & \cellcolor{Grey2} 0.158 & \cellcolor{Grey2} 0.24  & \cellcolor{Grey2} 2.89 & \cellcolor{Grey2} 43.186 & \cellcolor{Grey2} 1.84 & \cellcolor{Grey2} 0.23 & \cellcolor{Grey2} 3.52\\ \cline{2-13}

    & \multirow{3}{*}{LSTM*} & \cellcolor{White} N100 & \cellcolor{White} 3.66 & \cellcolor{White} 57.29 & \cellcolor{White} 3.13 & \cellcolor{White} 0.30 & \cellcolor{White} 5.74 & \cellcolor{White} 2.18 & \cellcolor{White} 31.22 & \cellcolor{White} 0.38 & \cellcolor{White} 0.40 & \cellcolor{White} 15.13 \\ \cline{3-13}
     & ~ & \cellcolor{Grey1} S\&P & \cellcolor{Grey1} {0.38} & \cellcolor{Grey1} {4.89} & \cellcolor{Grey1} -0.84 & \cellcolor{Grey1} {0.18} & \cellcolor{Grey1} {1.16} & \cellcolor{Grey1} {1.05} & \cellcolor{Grey1} {14.06} & \cellcolor{Grey1} 0.255 & \cellcolor{Grey1} 0.24 & \cellcolor{Grey1} 4.59 \\ \cline{3-13}
     & ~ & \cellcolor{Grey2} NIFTY & \cellcolor{Grey2} 2.08 & \cellcolor{Grey2} 29.61 & \cellcolor{Grey2} 1.81 & \cellcolor{Grey2} {0.15} & \cellcolor{Grey2} 0.62 & \cellcolor{Grey2} 2.64 & \cellcolor{Grey2} {38.86} & \cellcolor{Grey2} 1.24 & \cellcolor{Grey2} {0.23} & \cellcolor{Grey2} {2.18} \\ \cline{2-13}

    & \multirow{3}{*}{WOTPP} & \cellcolor{White} N100 & \cellcolor{White} {2.43} & \cellcolor{White} {35.32} & \cellcolor{White} {3.48} & \cellcolor{White} {0.28} & \cellcolor{White} {4.58} & \cellcolor{White} {1.42} & \cellcolor{White} 19.44& \cellcolor{White} {0.68} & \cellcolor{White} {0.379} & \cellcolor{White} {11.34}  \\ \cline{3-13}
     & ~ & \cellcolor{Grey1} S\&P & \cellcolor{Grey1} 1.95 & \cellcolor{Grey1} 27.54 & \cellcolor{Grey1} {3.20} & \cellcolor{Grey1} 0.182 & \cellcolor{Grey1} 2.20 & \cellcolor{Grey1} {1.40} & \cellcolor{Grey1} 19.14 &  \cellcolor{Grey1} 0.738 & \cellcolor{Grey1} 0.24 & \cellcolor{Grey1} {4.24} \\ \cline{3-13} 
     & ~ & \cellcolor{Grey2} NIFTY & \cellcolor{Grey2} 6.09 & \cellcolor{Grey2} 110.61& \cellcolor{Grey2} 6.96 & \cellcolor{Grey2} 0.16 & \cellcolor{Grey2} 1.66 &  \cellcolor{Grey2} {2.22} & \cellcolor{Grey2} 31.87 & \cellcolor{Grey2} 1.54 & \cellcolor{Grey2} 0.24 & \cellcolor{Grey2} 2.89 \\ 

     \hline
     \hline
\end{tabular}
\end{scriptsize}
\end{table*}

\begin{table*}
\caption{Performance of the TA-HKGE model with a 750-day training period (750TR) and a 1000-day training period (1000TR).}
\label{days_increase}
\begin{scriptsize}
\begin{center}
\begin{tabular}{l l l lllll | lllll }
    \hline
    \hline
        \multirow{2}{*}{$\Delta$} & \multirow{2}{*}{Model} & \multirow{2}{*}{Dataset} & \multicolumn{5}{c|}{Top 1} &\multicolumn{5}{c}{Top 5} \\ \cline{4-13}
         &  &  & IRR & AIRR & SR & NDCG & ACC & IRR & AIRR & SR & NDCG & ACC \\ 
         \hline
         \hline
         \multirow{6}{*}{1} & \multirow{3}{*}{750TR} & \cellcolor{White} N100 & \cellcolor{White} 0.112 & \cellcolor{White} 32.47 & \cellcolor{White} 0.652 & \cellcolor{White} 0.268 & \cellcolor{White} 2.5 & \cellcolor{White} 0.121 & \cellcolor{White} 35.67 & \cellcolor{White} 0.428 & \cellcolor{White} 0.375 & \cellcolor{White} 12.12 \\ \cline{3-13}
          &  & \cellcolor{Grey1} S\&P & \cellcolor{Grey1} 0.004 & \cellcolor{Grey1} 0.92 & \cellcolor{Grey1} 0.179 & \cellcolor{Grey1} 0.154 & \cellcolor{Grey1} 0.25 & \cellcolor{Grey1} 0.066 & \cellcolor{Grey1} 18.02 & \cellcolor{Grey1} 0.306 & \cellcolor{Grey1} 0.225 & \cellcolor{Grey1} 2.62 \\ \cline{3-13}
         &  & \cellcolor{Grey2} NIFTY & \cellcolor{Grey2} 0.139 & \cellcolor{Grey2} 41.83 & \cellcolor{Grey2} -499.2 & \cellcolor{Grey2} 0.163 & \cellcolor{Grey2} 0.57 & \cellcolor{Grey2} 0.094 & \cellcolor{Grey2} 26.7 & \cellcolor{Grey2} 0.470 & \cellcolor{Grey2} 0.233 & \cellcolor{Grey2} 1.97 \\ \cline{2-13}
         
         &  \multirow{3}{*}{1000TR} & \cellcolor{White} N100 & \cellcolor{White} 0.059 & \cellcolor{White} 16.18	& \cellcolor{White} 0.266 & \cellcolor{White} 0.271 & \cellcolor{White} 0.377 & \cellcolor{White} 0.123 & \cellcolor{White} 36.22 & \cellcolor{White} 0.448 & \cellcolor{White} 0.382 & \cellcolor{White} 11.58 \\ \cline{3-13}
         & ~ & \cellcolor{Grey1} S\&P & \cellcolor{Grey1} 0.158 & \cellcolor{Grey1} 48.78 & \cellcolor{Grey1} 0.463 & \cellcolor{Grey1} 0.162 & \cellcolor{Grey1} 1.16 & \cellcolor{Grey1} 0.078 & \cellcolor{Grey1} 21.81 & \cellcolor{Grey1} 0.329 & \cellcolor{Grey1} 0.233 & \cellcolor{Grey1} 2.15 \\ \cline{3-13}
         & ~ & \cellcolor{Grey2} NIFTY & \cellcolor{Grey2} 0.087 & \cellcolor{Grey2} 24.38 & \cellcolor{Grey2} 0.515 & \cellcolor{Grey2} 0.155 & \cellcolor{Grey2} 0.17 & \cellcolor{Grey2} 0.111 & \cellcolor{Grey2} 32.4 & \cellcolor{Grey2} 0.485 & \cellcolor{Grey2} 0.227 & \cellcolor{Grey2} 1.83 \\ \cline{1-13}

         \hline 
         \hline

        \multirow{6}{*}{5} & \multirow{3}{*}{750TR} & \cellcolor{White} N100 & \cellcolor{White} 0.38 & \cellcolor{White} 21.58 & \cellcolor{White} 1.043 & \cellcolor{White} 0.275 & \cellcolor{White} 2.49 & \cellcolor{White} 0.617 & \cellcolor{White} 36.34 & \cellcolor{White} 1.035 & \cellcolor{White} 0.384 & \cellcolor{White} 12.46 \\ \cline{3-13}
         & ~ & \cellcolor{Grey1} S\&P & \cellcolor{Grey1} 0.718 & \cellcolor{Grey1} 43.42 & \cellcolor{Grey1} 1.923 & \cellcolor{Grey1} 0.166 & \cellcolor{Grey1} 1.05 & \cellcolor{Grey1} 0.457 & \cellcolor{Grey1} 25.85 & \cellcolor{Grey1} 0.856 & \cellcolor{Grey1} 0.236 & \cellcolor{Grey1} 3.39 \\ \cline{3-13}
         & ~ & \cellcolor{Grey2} NIFTY & \cellcolor{Grey2} 0.447 & \cellcolor{Grey2} 25.2 & \cellcolor{Grey2} 1.008 & \cellcolor{Grey2} 0.159 & \cellcolor{Grey2} 0.5 & \cellcolor{Grey2} 0.332 & \cellcolor{Grey2} 18.21 & \cellcolor{Grey2} 0.665 & \cellcolor{Grey2} 0.231 & \cellcolor{Grey2} 1.93  \\ \cline{2-13}

         & \multirow{3}{*}{1000TR} & \cellcolor{White} N100 & \cellcolor{White} 0.411 & \cellcolor{White} 22.97 & \cellcolor{White} 0.984 & \cellcolor{White} 0.279 & \cellcolor{White} 3.27 & \cellcolor{White} 0.645 & \cellcolor{White} 38.25 & \cellcolor{White} 1.006 & \cellcolor{White} 0.390 & \cellcolor{White} 13.81  \\ \cline{3-13}
         & ~ & \cellcolor{Grey1} S\&P & \cellcolor{Grey1} 0.142 & \cellcolor{Grey1} 7.4 & \cellcolor{Grey1} 0.597 & \cellcolor{Grey1} 0.169 & \cellcolor{Grey1} 1.03 & \cellcolor{Grey1} 0.621 & \cellcolor{Grey1} 36.65 & \cellcolor{Grey1} 1.433 & \cellcolor{Grey1} 0.244 & \cellcolor{Grey1} 4.14 \\ \cline{3-13}
         & ~ & \cellcolor{Grey2} NIFTY & \cellcolor{Grey2} 0.901 & \cellcolor{Grey2} 57.1 & \cellcolor{Grey2} 2.102 & \cellcolor{Grey2} 0.157 & \cellcolor{Grey2} 0.71 & \cellcolor{Grey2} 0.638 & \cellcolor{Grey2} 37.76 & \cellcolor{Grey2} 1.181 & \cellcolor{Grey2} 0.23 & \cellcolor{Grey2} 2.16 \\ \cline{1-13}

         \hline
         \hline

        \multirow{6}{*}{20} & \multirow{3}{*}{750TR} & \cellcolor{White} N100 & \cellcolor{White} 0.378 & \cellcolor{White} 4.87 & \cellcolor{White} 1.536 & \cellcolor{White} 0.258 & \cellcolor{White} 2.17 & \cellcolor{White} 1.456 & \cellcolor{White} 19.98 & \cellcolor{White} 1.698 & \cellcolor{White} 0.3723 & \cellcolor{White} 11.58 \\ \cline{3-13}
         & ~ & \cellcolor{Grey1} S\&P & \cellcolor{Grey1} 0.588 & \cellcolor{Grey1} 7.66 & \cellcolor{Grey1} 1.254 & \cellcolor{Grey1} 0.187 & \cellcolor{Grey1} 1.35 & \cellcolor{Grey1} 1.256 & \cellcolor{Grey1} 17.03 & \cellcolor{Grey1} 1.420 & \cellcolor{Grey1} 0.258 & \cellcolor{Grey1} 4.34 \\ \cline{3-13}
         & ~ & \cellcolor{Grey2} NIFTY & \cellcolor{Grey2} 2.637 & \cellcolor{Grey2} 38.8 & \cellcolor{Grey2} 3.404 & \cellcolor{Grey2} 0.162 & \cellcolor{Grey2} 0.1 & \cellcolor{Grey2} 2.332 & \cellcolor{Grey2} 33.7 & \cellcolor{Grey2} 2.107 & \cellcolor{Grey2} 0.235 & \cellcolor{Grey2} 2.51  \\ \cline{2-13}

         & \multirow{3}{*}{1000TR} & \cellcolor{White} N100 & \cellcolor{White} 1.334 & \cellcolor{White} 18.18 & \cellcolor{White} 1.433 & \cellcolor{White} 0.265 & \cellcolor{White} 4.01 & \cellcolor{White} 2.082 & \cellcolor{White} 29.64 & \cellcolor{White} 2.440 & \cellcolor{White} 0.373 & \cellcolor{White} 11.35 \\ \cline{3-13}
         & ~ & \cellcolor{Grey1} S\&P & \cellcolor{Grey1} 0.907 & \cellcolor{Grey1} 12.05 & \cellcolor{Grey1} 1.767 & \cellcolor{Grey1} 0.162 & \cellcolor{Grey1} 0.11 & \cellcolor{Grey1} 1.853 & \cellcolor{Grey1} 26.02 & \cellcolor{Grey1} 1.945 & \cellcolor{Grey1} 0.237 & \cellcolor{Grey1} 3.6 \\ \cline{3-13}
         & ~ & \cellcolor{Grey2} NIFTY & \cellcolor{Grey2} 0.613 & \cellcolor{Grey2} 8.0 & \cellcolor{Grey2} 0.565 & \cellcolor{Grey2} 0.147 & \cellcolor{Grey2} 0 & \cellcolor{Grey2} 1.679 & \cellcolor{Grey2} 23.34 & \cellcolor{Grey2} 1.724 & \cellcolor{Grey2} 0.219 & \cellcolor{Grey2} 1.33    \\ \cline{1-13}

         \hline 
         \hline
\end{tabular}
\end{center}
\end{scriptsize}
\end{table*}

\section{Data Description}
\label{appdx:sec: datadesc}
The dataset consists of three main components present in the \texttt{Phase-Stock-KG/data/} directory:
\begin{itemize}
    \item \texttt{[country]\_relations.json} – Contains relationships between entities in the stock market of the USA or India.
    \item \texttt{[country]\_nodes.json} – Represents the nodes (companies, sectors) in the knowledge graph.
    \item \texttt{[company\_ticker]-[name].csv} – Contains historical stock price movements, including features Close, Open, Low, and High prices, and trading volume for each trading day.
\end{itemize}

\subsection{\texttt{[country]\_relations.json} \& \texttt{[country]\_nodes.json}}

These files contain the relationships between different entities in the stock market and nodes in the knowledge graph. The information required to create these files was derived from multiple sources, including:
\begin{itemize}
    \item \textbf{Financial Reports:} Balance sheets, cash flow statements, and financial statements of individual companies.
    \item \textbf{News Articles:} Relevant financial news that may influence stock prices and company relationships.
    \item \textbf{Events:} Events in India and the USA that could affect market trends.
\end{itemize}

\subsection{\texttt{[company\_ticker]-[name].csv}}
The directory \texttt{data/[nifty500/nasdaq100/sp500]} contains CSV files with time-series data of stock price indicators such as Open, High, Low, Close, and Volume for each company.

\subsection{Knowledge Graph Construction}
The \texttt{temporal\_kg\_nifty.pkl} and \texttt{temporal\_kg.pkl} files, for the Indian and US markets, respectively, are generated by processing relationships and nodes extracted from stock market data. Initially, JSON files such as \texttt{[country]\_relations.json} and \texttt{[country]\_nodes.json} are loaded to extract the relationships between different entities and the corresponding nodes in the knowledge graph. Each node is mapped to its corresponding attributes.

Once the nodes are mapped, relationships (consisting of a head entity, tail entity, relation type, and timestamps) are iterated and linked to the respective entities, forming a structured knowledge graph. The temporal aspect was introduced by incorporating time-based stock market events from CSV files containing financial statements, news articles, and event data from India and the USA. These time-sensitive events are mapped to the corresponding entities in the graph to ensure a dynamic representation of market fluctuations.

\begin{figure*}[ht]
  \centering
  \begin{subfigure}[t]{0.45\linewidth}
    \centering
    \includegraphics[width=\linewidth]{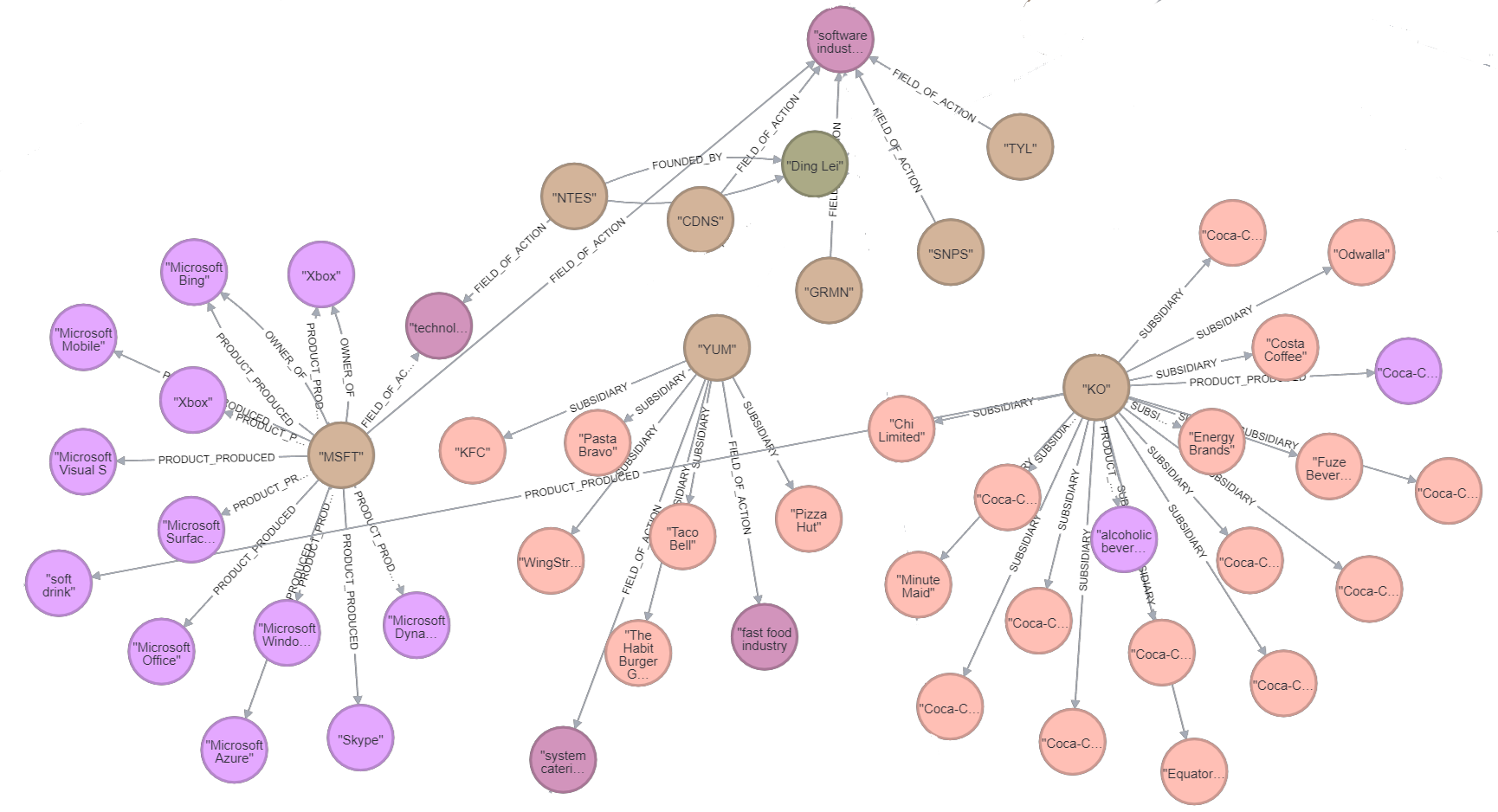}
    \caption{NASDAQ}
    \label{subfig:nasdaq_graph}
  \end{subfigure}
  \hfill
  \begin{subfigure}[t]{0.45\linewidth}
    \centering
    \includegraphics[width=\linewidth]{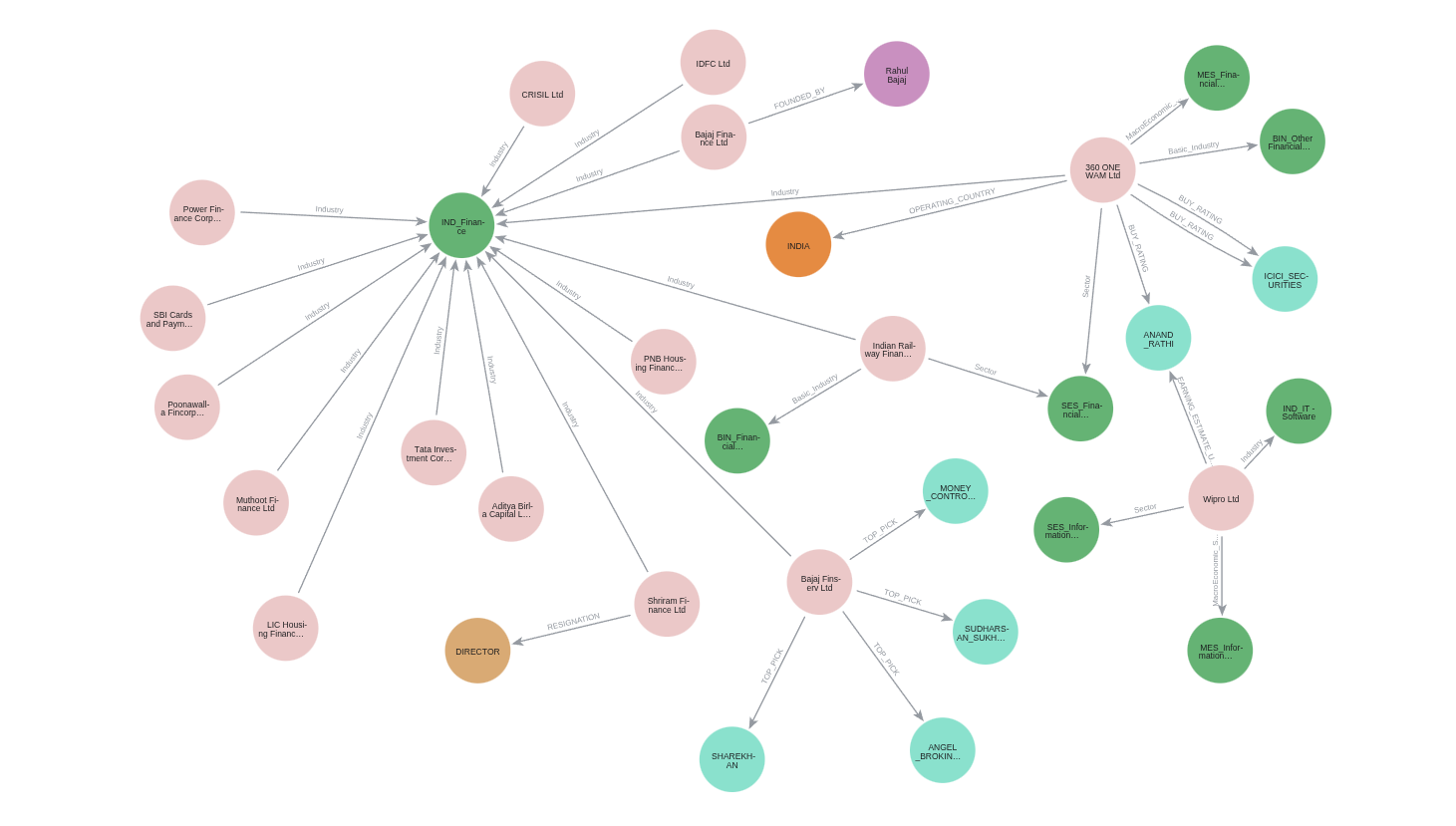}
    \caption{NSE}
    \label{subfig:nse_graph}
  \end{subfigure}
  \caption{Visualization of a subgraph from the constructed $\mathsf{STOCKnowledge}$s for (left) NASDAQ and (right) NSE share markets. Entities and relations are highlighted by type.}
  \label{Figure: KG diagrams}
\end{figure*}
\paragraph{Handling timestamps:}
\begin{itemize}
\item If an expiry timestamp is available, it is used.
\item If only a timestamp is present, expiry is assumed as one day later.
\item If no timestamp exists, it defaults to 1970-01-01.
\end{itemize}
The processed relations are stored as
\begin{equation}
[\; head,\; tail,\; relation\_id,\; timestamp,\; expiry\_timestamp\;].
\end{equation}
The data were sorted by timestamp for chronological ordering. Finally, the processed temporal knowledge graph was serialized and stored as \texttt{temporal\_kg\_nifty.pkl} and \texttt{temporal\_kg.pkl} using the pickle. 

\subsection{Batch Dataset Creation from Historical CSV and PKL Files}
The batch dataset was constructed using historical stock data from CSV files and a temporal knowledge graph from PKL files.

\subsubsection{Stock Data from CSV Files}
\begin{itemize}
    \item Iterate over stock price CSV files stored in the directory (e.g., \texttt{data/nasdaq100/}).
    \item Each file corresponds to a company and contains columns: Date, Open, High, Low, Close, Volume.
    \item Ignore files with less than $2800$ rows to ensure sufficient training samples.
\end{itemize}

\subsubsection{Time-Series Windowing}
For each company:
\begin{itemize}
    \item Apply a sliding window approach with window size $W$ and prediction horizon $\Delta$.
    \item Normalize input features by dividing by the first value in the window.
    \item Predict returns at different horizons: $T = \{1, 5, 20\}$.
    \item Extract low and high prices for multiple horizons.
\end{itemize}

\subsubsection{Loading the temporal knowledge graph}
\begin{itemize}
    \item Load the preprocessed knowledge graph from \texttt{temporal\_kg.pkl}.
    \item Extract relevant temporal facts corresponding to each time window.
    \item Constructs a sector-based graph using adjacency relations.
\end{itemize}

\subsubsection{Final Dataset Construction}
For each time window:
\begin{itemize}
    \item Collect stock price window features and targets.
    \item Extract corresponding temporal KG snapshot.
    \item Form a dataset tuple:
    \begin{equation}
        (\text{Stock Price Window Features}, \text{Temporal KG})
    \end{equation}
\end{itemize}

The final output is a file \texttt{data/pickle/[dataset]/full\_graph\_data.pkl} containing a dictionary with keys \texttt{train} and \texttt{company}, which consist of a batch dataset containing stock price sequences and the temporal knowledge graph and a dictionary of company IDs, respectively.

\subsection{Node Data Format}
The nodes JSON file consists of structured data that represent entities in the knowledge graph. Each node entry contains the following:
\begin{itemize}
    \item \textbf{identity}: A unique identifier for the node.
    \item \textbf{labels}: A list of labels describing the type of the entity (e.g., "Person", "Company", etc.).
    \item \textbf{properties}: Key-value pairs defining attributes of the node, such as \texttt{name} and \texttt{id}
    \item \textbf{elementId}: A system-generated unique identifier for the node.
\end{itemize}

\noindent \textbf{Example JSON node entry:}
\begin{lstlisting}[language=json,frame=single]
{
  "n": {
    "identity": 0,
    "labels": ["Person"],
    "properties": {
      "name": "Tommy Millner",
      "id": 114689399
    },
    "elementId": "0"
  }
}
\end{lstlisting}

In this example, a node labeled Person represents Tommy Millner with an assigned ID of 114689399.

\subsection{Relation Data Format}
In this example, there is a "SUBSIDIARY" relationship between the nodes with identities 1007 and 2591, with a relationship ID of "P355".

The relations JSON file captures the relationships between nodes and describes how entities are linked. Each relationship entry consists of the following:
\begin{itemize}
    \item \textbf{identity}: A unique identifier for the relationship.
    \item \textbf{start}: The identity of the starting node in the relationship.
    \item \textbf{end}: The identity of the ending node in the relationship.
    \item \textbf{type}: A string indicating the relationship type (e.g., "SUBSIDIARY", "OWNS", "EMPLOYS", etc.).
    \item \textbf{properties}: Additional metadata, such as a unique id for the relationship.
    \item \textbf{elementId}: A system-generated identifier for the relationship.
    \item \textbf{startNodeElementId} / \textbf{endNodeElementId}: Unique element IDs of the connected nodes.
\end{itemize}

\noindent \textbf{Example JSON relation entry:}
\begin{lstlisting}[language=json,frame=single]
{
  "r": {
    "identity": 0,
    "start": 1007,
    "end": 2591,
    "type": "SUBSIDIARY",
    "properties": {
      "id": "P355"
    },
    "elementId": "0",
    "startNodeElementId": "1007",
    "endNodeElementId": "2591"
  }
}
\end{lstlisting}

This structured data enables efficient querying and analysis within the knowledge graph, forming the basis for graph-based learning and reasoning. The relationships between entities allow the construction of a knowledge-driven system for applications such as recommendation systems, risk analysis, and trend forecasting, among others.

\section{Web Scraping Details}
\label{appdx:sec: webscrap}
\subsection{Entity Recognition and Event Extraction for News Articles}

\paragraph{Entity Recognition}
Entities such as companies, dates, and financial metrics are recognized using predefined patterns and keywords in the news content. The functions earning\_pattern and guidance\_pattern use regular expressions and keyword matching to identify the relevant entities and events.

\paragraph{Event Extraction}
\begin{itemize}
    \item \textbf{\textit{Earnings Events}}: The earning\_pattern function looks for keywords like 'beat', 'miss', and 'in-line' within the news content to determine if a company has met, exceeded, or fallen short of earnings expectations. It extracts the relevant information and categorizes the event as "BEATS\_EXPECTATION", "MISSES\_EXPECTATION", or "INLINE\_EXPECTATION".
    \item \textbf{\textit{Guidance Events}}: The guidance\_pattern function searches for keywords indicating changes in company guidance, such as 'raise', 'cut', or 'in-line'. It categorizes the event as "RAISES\_GUIDANCE", "CUTS\_GUIDANCE", or "INLINE\_GUIDANCE".
\end{itemize}

\paragraph{Timestamp Extraction}
The get\_timestamp function extracts timestamps from the news content. It identifies the location of the timestamp within the content structure and converts it to a datetime object using the Pandas library.

\paragraph{Head and Tail Recognition}
\begin{itemize}
    \item \textbf{\textit{Head}}: The head of a triple typically represents the subject of the event, such as the company name. It was extracted from the news content using keyword matching and context analysis.
    \item \textbf{\textit{Tail}}: The tail of a triple represents the object or outcome of the event, such as the financial metric or result. It was identified using similar keyword matching and context analysis techniques.
\end{itemize}

This process ensures that financial events are accurately extracted and structured for further analysis or integrated into a graph database.

\subsection{Profile and Relationship}

The notebook \texttt{wikidata\_create.ipynb} was designed to create a knowledge graph (KG) from Wikidata for companies listed on NASDAQ and NYSE. It involves several steps, including data loading, processing, mapping, and storing the relationships between the companies and their attributes.

CSV files containing mappings between ticker symbols and Wikidata codes for NASDAQ and NYSE companies were loaded.

\subsubsection{Ticker List Creation}
Iterates through directories containing company data files to create a list of tickers with sufficient data points (more than 2600 rows).

\subsubsection{Wikidata Querying}
Using pywikibot, the notebook queries Wikidata to fetch the relationships and attributes of the companies. Relationships such as "OWNED\_BY", "SUBSIDIARY", "PARENT\_ORGANISATION", "CEO", etc., are considered.

\section{Time Complexity}
\label{appdx:timecomplexity}
We analyze the computational complexity of the TA-HKGE in terms of the key problem dimensions.
Let $N$ be the number of stock assets, $W$ the historical window length, $d_s$ the sequential embedding dimension, $d_r$ the relational embedding dimension, $L$ the number of Transformer encoder layers, and $H$ the number of attention heads.
Let $|\mathcal{E}_T|$ denote the number of temporal KG edges (events) in a snapshot at time $T$, $|\mathcal{E}_0|$ the number of static KG edges, $L$ the number of Transformer encoder layers, and $H$ the number of attention heads.
Let $d_{tpp}$ be the temporal point-process embedding dimension used by the HPGE component, and $K$ be the number of negative samples per event.
Let $|\mathcal{E}_T|$ be the number of temporal KG edges (events) in a snapshot at time $T$, and $|\mathcal{E}_0|$ is the number of static KG edges.
Let $M$ be the number of relation types (edge types) in a heterogeneous KG.

\textbf{Sequential encoder (Transformer)}
For a standard full-attention Transformer encoder over a length-$W$ sequence, the per-layer time complexity is $\mathcal{O}(W^2 d_s + W d_s^2)$.
Thus, computing the sequential embeddings for all $N$ assets costs
$\mathcal{O}\big(N L (W^2 d_s + W d_s^2)\big)$.

\textbf{Temporal process embeddings (HPGE/Hawkes on graphs)}
Learning temporal process embeddings scales with the number of observed events and sampled negatives.
If we process $|\mathcal{E}_T|$ temporal events and sample $K$ negatives per event, the dominant term is
$\mathcal{O}(|\mathcal{E}_T|\,K\,d_{tpp})$ for embedding updates (with additional overhead from neighborhood aggregation depending on the selected history truncation).

\subsection{Relational encoder (heterogeneous attention over KG)}
For a message-passing layer with attention over edges, the complexity is linear in the number of edges times the embedding dimension.
With $L_g$ HEAT layers (we use $L_g=2$), the per-snapshot complexity is approximately
$\mathcal{O}\big( L_g(|\mathcal{E}_0|+|\mathcal{E}_T|)\,d_r\big)$,
ignoring constant factors from relation-type-specific projections.

\subsection{Prediction and loss}
The final ranking layer is $\mathcal{O}(N d)$ where $d(d_s+d_{tpp}+d_r)$ is the concatenated embedding dimension.
The listwise ApproxNDCG loss requires pairwise comparisons within a batch; if computed over all $N$ assets at a time point, it contributes $\mathcal{O}(N^2)$ per time step.

\textbf{Combined complexity} For one time point $T$ (on one snapshot), an end-to-end training step can be summarized as
\begin{align*}
\mathcal{O}&\Big( NL(W^2 d_s + W d_s^2)
+ |\mathcal{E}_T|K d_{tpp}
+ L_g(|\mathcal{E}_0|+|\mathcal{E}_T|)d_r + \mathcal{O}(N d) + \mathcal{O}(N^2)
\Big),
\end{align*}

\end{document}